\documentclass[times,final]{elsarticle}

\usepackage[margin=1in]{geometry}
\def\snm#1{#1}
\newcommand{\verso}[1]{}

\usepackage{framed,multirow}

\usepackage{amssymb}
\usepackage{latexsym}

\usepackage[ruled,vlined]{algorithm2e}
\usepackage{physics}
\usepackage{bm}
\usepackage{subcaption}
\usepackage{bbold}
\usepackage{array}
\usepackage{mathtools}

\DeclareUnicodeCharacter{2212}{−}
\usepackage{url}
\usepackage{xcolor}
\usepackage{pgfplots}
\usepgfplotslibrary{groupplots,dateplot}
\pgfplotsset{compat=newest}
\usepackage{tikz}
\usetikzlibrary{patterns,shapes.arrows,spy,external,math}

\definecolor{crimson2143940}{RGB}{214,39,40}
\definecolor{darkgray176}{RGB}{176,176,176}
\definecolor{darkorange25512714}{RGB}{255,127,14}
\definecolor{forestgreen4416044}{RGB}{44,160,44}
\definecolor{lightgray204}{RGB}{204,204,204}
\definecolor{steelblue31119180}{RGB}{31,119,180}

\usepackage[normalem]{ulem}
\definecolor{readable_green}{RGB}{40, 189, 16}

\newcommand{\ReviewerOne}[1]{#1}
\newcommand{\ReviewerTwo}[1]{#1}
\newcommand{\ReviewerThree}[1]{#1}
\newcommand{\ReviewerOneStrike}[1]{}
\newcommand{\ReviewerTwoStrike}[1]{}
\newcommand{\ReviewerThreeStrike}[1]{}

\journal{Journal of Computational Physics}

\begin{document}

\verso{Xinjie Ji \textit{et al.}}

\begin{frontmatter}

\title{A sharp immersed method for 2D flow-body interactions using the vorticity-velocity Navier-Stokes equations}

\author[1]{Xinjie  \snm{Ji}\fnref{fn1}}
\author[1]{James  \snm{Gabbard}\fnref{fn1}}
\author[1]{Wim M. \snm{van Rees}\corref{cor1}}
\cortext[cor1]{Corresponding author. E-mail address: wvanrees@mit.edu  
}

\address[1]{Department of Mechanical Engineering, Massachusetts Institute of Technology, 77 Massachusetts Avenue, Cambridge, MA 02139, USA} %
\fntext[fn1]{Contributed equally}

\begin{abstract}

\noindent Immersed methods discretize boundary conditions for complex geometries on background Cartesian grids. This makes such methods especially suitable for two-way coupled flow-body problems, where the body mechanics are partially driven by hydrodynamic forces. However, for the vorticity-velocity form of the Navier-Stokes equations, existing immersed geometry discretizations for two-way coupled problems only achieve first order spatial accuracy near solid boundaries. Here we introduce a sharp-interface approach based on the immersed interface method to handle \ReviewerOneStrike{one- and two-way coupled flow-body problems for} \ReviewerOne{the one- and two-way coupling between an incompressible flow and one or more rigid bodies using} the 2D vorticity-velocity Navier-Stokes equations. Our main contributions are three-fold. First, we develop and analyze a moving boundary treatment for sharp immersed methods \ReviewerTwo{that can be applied to PDEs with
}implicitly defined boundary conditions, such as those commonly imposed on the vorticity field. Second, we develop a two-way coupling methodology for the vorticity-velocity Navier-Stokes equations based on control-volume momentum balance that does not require the pressure field. Third, we show through extensive testing and validation that our resulting flow-body solver reaches second-order accuracy for most practical scenarios, and provides significant efficiency benefits compared to a representative first-order approach.

\end{abstract}

\end{frontmatter}

\section{Introduction}

Fluid-structure interaction problems are typically governed by complex geometries undergoing non-trivial, large, and \textit{a priori} unknown motions. In this context, immersed geometry methods provide an attractive alternative to traditional body-fitted grid approaches because they allow geometries to move and/or deform independently of the mesh \ReviewerOne{\cite{griffith2020immersed, mittal2005}}. For \ReviewerOne{2D} external flows it is appealing to combine immersed methods with the vorticity-velocity formulation of the Navier-Stokes equation, \ReviewerOne{which can reduce simulation costs through the use of compact computational domains that only contain the support of the vorticity field. This formulation also eliminates much of the numerical complexity associated with the pressure field, such as specialized time integration schemes to enforce the incompressibility constraint or odd-even decoupling in the pressure Poisson equation on collocated grids.} However, to the best of our knowledge no immersed geometry methodology currently exists for 2D vorticity-velocity flow solvers that can handle two-way coupled flow-body problems with spatio-temporal accuracy beyond first-order.\footnote{We use two-way coupling to denote that the body motion is partially driven by the flow forces, rather than being fully imposed.} To provide further context to this open challenge, below we will present a brief survey of (1) existing embedded boundary approaches for 2D vorticity-velocity flow solvers, and (2) the treatment of moving boundaries in sharp embedded boundary methods.

There are broadly three categories of embedded boundary approaches for 2D vorticity-velocity Navier-Stokes solvers, where we exclude classical panel-method type approaches \cite{Eldredge2007, Eldredge2008} that require the formulation and solution of a boundary integral problem. The first category contains discrete-forcing approaches based on the immersed interface method (IIM). In \cite{Calhoun2002,Li2003}, the first IIM simulations of flows past single, stationary embedded bodies were reported, achieving up to second-order convergence in space and time. This was extended to simulations past multiple, moving bodies \cite{Russell2003} using an overset grid to compute boundary vorticity values, reducing the convergence of the near-wall vorticity field to first order. For single, stationary bodies an explicit-jump immersed interface method was used in \cite{Linnick2005}, where the use of compact finite-difference schemes led to fourth-order convergence. A similar approach was recently used in \cite{Singhal2022}, where flows past multiple, moving bodies were considered using compact, fourth-order spatially accurate finite-difference schemes. Immersed interface methods have also been used in combination with a vortex-particle method, first using a Lighthill-type splitting method \cite{Marichal2016,MarichalThesis} and subsequently using a Thom-like vorticity boundary condition \cite{Gillis2019}, both achieving second order accuracy. Finally, in a previous paper \cite{Gabbard2021}, we proposed a fully explicit second-order accurate IIM-based vorticity-velocity Navier-Stokes solver that handle multiple stationary bodies using conservative finite-difference methods. The second and third categories contain continuous-forcing approaches based on the penalization method and immersed boundary method (IBM), respectively. The penalization method was proposed in \cite{Angot:1999} and combined with remeshed vortex methods in \cite{Coquerelle:2008}. Subsequently, this approach has been developed and enhanced over a sequence of works to handle multiple moving and deforming bodies as well as two-way coupling problems \cite{Rossinelli2010,Gazzola:2011a,Rossinelli2015,Gillis2018,Bernier2019,Bhosale2021}, all while retaining fundamentally a first-order accuracy in space. \ReviewerTwo{\citet{engels2015numerical} proposed a similar approach that combines the penalization method with a pseudospectral spatial discretization for simulations with thin solid bodies and two-way coupled fluid-structure interaction.} Lastly, an immersed boundary method for the vorticity-velocity formulation was first proposed in \citep{Poncet2009}, \ReviewerTwo{and subsequent IBMs have been developed for flows with heat transfer \cite{Soltani2016} and for simulations based on a meshless spatial discretization \cite{bourantas2023immersed}}. More recently, \citet{WandE2015} presented a projection immersed boundary method for two-way coupling problems with a strong-coupling scheme. The strong-coupling method enables stable simulation of bodies with an extremely small density ratio, but the convergence rate of spatial accuracy is still first-order near the boundary. Surveying all these contributions, we conclude that in the context of vorticity-velocity formulation with embedded moving boundaries, only the IIM provides a sharp boundary discretization. Consequently, only the IIM-based approaches have yielded spatial convergence rates beyond first order in the infinite norm. None of these IIM methods, however, have been extended to two-way coupling problems. One fundamental challenge related to this is that the vorticity-velocity formulation does not explicitly yield a pressure field. For the IBM or penalization method, this can be circumvented because the fictitious forces employed to enforce boundary conditions on the flow can be directly related to the forces applied by the flow on the body. For most sharp IIM approaches this approach is unavailable and body forces are typically obtained through control volume analyses, which are more challenging to extend to coupled flow-body problems.

Nevertheless, sharp immersed methods are particularly advantageous for simulations with moving boundaries, providing accuracy on immersed interfaces while avoiding the cost and complexity of maintaining a moving body-fitted mesh. In order to achieve second-order spatial accuracy and higher for moving boundaries, sharp immersed methods must deal with the challenge of ``freshly cleared cells" that move from one side of an interface to the other during the course of a time step. This crossing generally creates a discontinuity in the time history of the cell that must be explicitly addressed to maintain accuracy near the interface. As noted by \cite{Udaykumar1999}, this issue is unique to sharp immersed methods with moving boundaries and does not appear when using a body fitted mesh, a continuous-forcing immersed method, or a sharp immersed methods with stationary boundaries. 

One of the earliest methods for treating freshly cleared cells appears in the immersed interface method of \citet{Li1997}, which discretizes the Navier-Stokes equations with moving boundaries using a semi-implicit Adams-Bashforth/Crank-Nicolson (ABCN) time integration scheme. The authors solve a system of equations to determine the time at which each uncovered point crosses the immersed interface, as well as the jumps in the solution and its temporal derivative at the crossing point. All of this information is used to modify the ABCN integration scheme at the uncovered point to maintain first order temporal accuracy. \citet{Brehm2011} further develop this method to allow for full second order temporal accuracy near the immersed interface. Their method uses a Crank-Nicolson scheme along with a specialized semi-implicit discretization of the nonlinear convective term and relies on a no-slip boundary condition to relate derivatives of the boundary motion to derivatives of the unknown solution. A similar approach was pursued by \citet{Xu2006}, who found that incorporating a temporal jump condition in each stage of a fourth order Runge-Kutta scheme yields between first and second order temporal accuracy in the $L_\infty$ norm, and only slightly reduces error when compared compared to an approach that fully neglects the jumps. While these approaches successfully allow for sharp moving interfaces, they are all specialized to a particular integration scheme and are difficult to extend to higher order accuracy or PDEs with more complex boundary conditions.

A more general and widely successful approach to treating uncovered points was developed in \cite{Udaykumar1999} for diffusion-driven interface growth problems and in \cite{Udaykumar2001} for the 2D incompressible Navier-Stokes equations. These methods determine the value of the solution at uncovered points from a second-order spatial interpolation that uses neighboring solution values and a prescribed boundary condition on the interface. This strategy avoids the need to explicitly identify crossing times and generalizes easily to other semi-implicit time integration schemes. For the incompressible Navier-Stokes equations the interpolation-based method was extended to 3D simulations of biological propulsion mechanisms in \cite{Mittal2008}. %
For the compressible Navier-Stokes equations, a similar interpolation-based methodology has been extended to higher-order interpolants and higher-order RK integrators for use in large-scale 3D aeroacoustics simulations \cite{Brehm2019} \ReviewerOne{and in 3D simulations with thin compliant structures \cite{boustani2021immersed}}. A variation on the interpolation method is introduced in \cite{Yang2006}, which allows for moving boundaries and a fully explicit three-stage RK scheme. In this field extension approach the velocity and pressure fields are extrapolated to points which lie inside a moving solid domain at the end of each time step. Additionally, during the time step when these solid points are uncovered their value is interpolated from surrounding solution values and a nearby boundary condition. This combination of field extensions and interpolation ensures that points entering the domain have a value at the current and previous time steps, providing a time history that can be used by multistep time integrators. The authors demonstrate that this approach produces second order accuracy in the $L_1$ and $L_2$ norms, and successfully apply their scheme to LES simulations of turbulent flows in complex geometries. While the field extension approach of \citet{Yang2006} succeeds in providing a time history of the solution at uncovered points, the interpolation step used in this method requires a known boundary condition at each time step. This is not the case for vorticity-velocity formulations of the 2D Navier-Stokes equations, in which the boundary condition on the vorticity depends on the full vorticity field and is not available at the start of each time step \cite{Gillis2019,Gabbard2021}.

In the current state-of-the-art on sharp-interface vorticity-velocity solvers, we can thus identify two open challenges: the enforcement of vorticity-based boundary conditions on moving sharp interfaces, and devising two-way coupling algorithms that do not rely on a pressure field. In this work we aim to address the above challenges to develop an IIM-based vorticity-velocity solver that can handle moving boundaries and two-way coupled problems \ReviewerOne{of one or more rigid bodies}. In section~\ref{section:IIM} we describe a variation on the field extension approach for moving sharp boundaries, that replaces the interpolation step with an extrapolation of the time derivative. The resulting method has several attractive properties that extend to PDEs other than the vorticity-velocity Navier-Stokes equations, including compatibility with higher-order RK schemes and a simplified implementation that does not require the identification of uncovered points. Through careful numerical experimentation we demonstrate that the additional errors introduced by this moving boundary treatment do not dominate the existing spatial and temporal errors for simulations in which the time step is limited by a CFL constraint. In section~\ref{section:algorithm} we combine this approach with our existing second-order Navier-Stokes solver for stationary bodies \cite{Gabbard2021}, and \ReviewerTwo{show its validation and convergence analyses of the 2D Navier-Stokes solver. In section~\ref{section:NS_twoWay} we present our extension towards two-way coupled flow-body problems of rigid bodies and validate the momentum balance algorithm through different test cases.} \ReviewerTwoStrike{present our extension towards two-way coupled flow-body problems. Sections~\ref{section:results_oneWay} and \ref{section:results_twoWay} show extensive validation and convergence analyses of the 2D Navier-Stokes solver for one-way and two-way coupled problems respectively.} We compare our results with analytic solutions, results from literature, and results from an in-house penalization-based approach. Finally, in section~\ref{section:conclusion} we conclude our results and lay out thoughts for future research.

\section{Immersed interface method}\label{section:IIM}
In this section we introduce the immersed interface method (IIM) for irregular domains with fixed boundaries, then extend this method to moving boundaries while maintaining high-order temporal accuracy.

\subsection{IIM for fixed boundaries}\label{subsection:IIM_fixed}
In the IIM an irregular domain boundary is superimposed on a background Cartesian grid, and standard finite difference stencils are used for PDE discretizations away from the domain boundaries. Near the irregular boundary these stencils are no longer valid and require local corrections to retain their order of accuracy. This is accomplished by extending each field beyond the domain boundary using a high-order polynomial extrapolation and applying standard finite difference stencils to the extended field. Provided that the order of the extrapolation is sufficiently large, the order of accuracy of the original finite difference scheme is preserved.  

For illustration, consider a uniform 1D grid with grid spacing $h$ and grid points $x_j = jh$ for $j \in \mathbb{Z}$, as well as an irregular domain boundary at $x_\alpha$ with \ReviewerOne{$x_{i-1} < x_\alpha < x_{i}$} (Figure \ref{fig:extension_diagrams}). The position of the boundary is characterized by the non-dimensional distance $\psi = \abs{\ReviewerOne{x_{i}} - x_\alpha}/h$, which satisfies $0 < \psi < 1$. On this grid any smooth function $f(x)$ can be represented by the point values $f_j = f(x_j)$, possibly augmented by a prescribed Dirichlet boundary condition $f_\alpha = f(x_\alpha)$. To calculate the second derivative of $f(x)$, the second order difference stencil
\begin{equation}\label{eq:centered_example_stencil}
    \qty(\dv[2]{f}{x})_i \approx \frac{f_{i+1} - 2f_i + f_{i-1}}{h^2}
\end{equation}
can be applied to each grid point in the domain, which requires the known function values $\ReviewerOne{f_{i}, \,f_{i+1}}, \dots$ as well as a ghost value \ReviewerOne{$f_{i-1}$} which lies outside of the problem domain. To construct this ghost value, we distinguish two separate types of IIM field extensions: those with a boundary condition and those without. For an $N$-th order extension with boundary condition, an interpolating polynomial $p_N(x)$ of degree $N - 1$ is constructed using data at the $N$ points $\{x_\alpha, \ReviewerOne{x_{i+1}}, \,\dots,\, \ReviewerOne{x_{i+N-1}}\}$. The grid point \ReviewerOne{$x_{i}$} is omitted from the interpolation stencil to ensure that the interpolation is well-conditioned for arbitrarily small $\psi$. The interpolating polynomial is then used to define a ghost value $\ReviewerOne{f_{i-1}} = p_N(\ReviewerOne{x_{i-1}})$ at the point $x_i$. For an extension without boundary condition, the procedure is repeated using data at the points \ReviewerOne{$\{x_{i}, x_{i+1}, \,\dots, \,x_{i+N-1}\}$}, which excludes any boundary condition prescribed on the domain boundary. \ReviewerTwo{For larger finite difference stencils, additional ghost values $\{f_{i-2}, f_{i-3},\,\dots\}$ are constructed by evaluating the interpolating polynomial $p_N(x)$ at the corresponding points $\{x_{i-2}, x_{i-3},\,\dots\}$.} Generally an $M$-th order discretization of a $D$-th order differential operator will retain its accuracy provided that $N \ge M + D$; for this example, the stencil in (\ref{eq:centered_example_stencil}) will retain its second order accuracy near the boundary provided that $N \ge 4$.

This extension-based IIM extends well to complex 2D domains, and can be used to treat PDEs with Neumann boundary conditions \cite{MarichalThesis}, elliptic PDEs in unbounded domains \cite{Gillis2018}, and conservative finite difference schemes for advection-diffusion equations \cite{Gabbard2021}. Here we use the 2D extension methodology outlined in \citet{Gabbard2021}, which relies on 1D extrapolation along grid lines for computational efficiency and permits complex 2D domains with non-convex boundaries (Figure \ref{fig:extension_2d}). For brevity we omit the specifics of the extrapolation procedure here, focusing instead on the extension of a generic extrapolation-based IIM to time-dependent PDEs with moving boundaries.

\begin{figure}[htb]
    \centering
    \begin{subfigure}{0.45\textwidth}
        \centering
        \resizebox{\textwidth}{!}{\input{tikzfigures/extension_diagram}}
        \caption{1D extension.}
        \label{fig:extension_diagrams}
    \end{subfigure}
    \hspace{0.1\textwidth}
    \begin{subfigure}{0.3\textwidth}
        \centering
        \resizebox{\textwidth}{!}{\input{tikzfigures/extension_2d}}
        \caption{2D extension.}
        \label{fig:extension_2d}
    \end{subfigure}
    \caption[]{(a) One dimensional IIM extensions of order four both with 
    (\tikz[baseline] \draw[forestgreen4416044, very thick] (0pt, 2pt) -- (7pt, 2pt);) 
    and without 
    (\tikz[baseline] \draw[steelblue31119180, very thick] (0pt, 2pt) -- (7pt, 2pt);)
    a boundary condition. Both cases require four interpolation points to construct the extension value 
    (\tikz[baseline] \draw[very thick] (0.25pt, 0.25pt) rectangle (3.75pt, 3.75pt);), 
    but differ in whether to include the boundary condition at $x_\alpha$ 
    (\tikz[baseline] \draw[very thick] (2pt, 2pt) circle (2pt);) 
    or the field value at $\ReviewerOne{x_{i}}$ 
    (\tikz[baseline] \filldraw[very thick] (2pt, 2pt) circle (2pt);). 
    (b) Stencils used for a third order 2D extrapolation. Point (i) has only one neighbor in the problem domain, and consequently receives an extrapolation along only one coordinate direction. Point (ii) has two neighbors, so its value is the average of two separate one-dimensional extrapolations. Points (iii) and (iv) have two neighbors in the problem domain, but there are not enough points in the domain to allow for a third-order horizontal extrapolation stencil; consequently, each is filled from a vertical extrapolation only.
     } 
    \label{fig:extensions}
\end{figure}

\subsection{IIM for moving boundaries} \label{subsection:IIM_moving}

For time-dependent PDEs with stationary boundaries, the IIM can be combined with a method-of-lines time discretization in a straightforward way. The PDE solution is represented by its values on the Cartesian grid while spatial derivatives are discretized with the IIM, yielding a large system of ODEs that can be solved with standard time integrators. The picture is more complicated for PDEs with moving boundaries, primarily because grid points enter and exit the problem domain as the simulation progresses. Exiting points are typically easy to deal with: once a point leaves the domain it is no longer used in the spatial discretization, and the associated unknowns can be removed from the system of ODEs. Points which enter the domain (typically called ``fresh" or ``freshly cleared" cells \cite{Udaykumar1999, Udaykumar2001, Mittal2008, Seo2011, Brehm2011}) must be provided with an initial value before they are added to the system, and for more complex time integrators these points may also need some form of artificial time history. 

In this paper we investigate the issue of freshly cut cells in the context of multi-stage time integrators, particularly low-storage Runge-Kutta methods. For a generic discretized PDE of the form $\partial_t u = f(u, t)$ these integrators store only the state $u(x_j, t)$ and a single history field $y(x_j, t)$, which both reduces memory requirements and simplifies the task of providing a time history for freshly cleared cells. For an $s$-stage RK integrator integrator let $u^{(i)}$, $y^{(i)}$, and $t^{(i)}$ refer to the state, history, and simulation time at $i$-th stage. To advance the system from time $t^n$ to time $t^{n+1} = t^n + \Delta t$, the state and temporary register are initialized with $u^{(0)} = u^n$ and $y^{(0)} = 0$, and for $i = 1$ to $s$ the stage values are updated with the low-storage update
\begin{equation}
    \begin{aligned}
        y^{(i+1)} &= a_i y^{(i)} + \Delta t f(u^{(i)}, t^{(i)}) \\
        u^{(i+1)} &= u^{(i)} + b_i y^{(i)} \\
        t^{(i+1)} &= t^{(i)} + c_i \Delta t \\
    \end{aligned}
\end{equation}
The value of the next \ReviewerThree{step} is taken to be $u^{n+1} = u^{(s)}$. To apply a low-storage Runge-Kutta method to an IIM discretization with moving boundaries, both the state $u^{(i)}$ and the right hand side $f(u^{(i)}, t^{(i)})$ are extrapolated before applying each stage update, and all unused values that fall outside the domain at the next stage are removed after the update. To make this precise, \ReviewerThree{let $\Omega(t)$ represent the moving problem domain. We define} an operator $E_f^{(i)}[\cdot]$ that performs an IIM extrapolation without boundary condition (Figure \ref{fig:extension_diagrams}, %
blue) for a field defined on $\Omega(t^{(i)})$, constructing ghost values and storing them in locations outside of the problem domain. The operator $E_u^{(i)}[\cdot]$ is defined similarly, but performs an IIM extrapolation using any boundary conditions prescribed on the field $u$ (Figure \ref{fig:extension_diagrams},
green) . Finally, we define a zeroing operator $Z^{(i)}$ that sets all points that fall outside the domain $\Omega(t^{(i)})$ to a zero value. With this notation the IIM-adapted low-storage update for PDEs with moving boundaries becomes
\begin{equation}\label{eq:lsrk_moving}
    \begin{aligned}
        y^{(i+1)} &= a_i y^{(i)} + \Delta t E_f^{(i)} \qty[f(u^{(i)}, t^{(i)})] \\
        u^{(i+1)} &= Z^{(i+1)} \qty[E_u^{(i)}\qty[u^{(i)}] + b_i y^{(i)}] \\
        t^{(i+1)} &= t^{(i)} + c_i \Delta t \\
    \end{aligned}
\end{equation}
\ReviewerOne{In this work we choose extrapolation operators that act on all points with a nearest neighbor in the problem domain. Consequently, each point that enters the problem domain will have a valid time history so long as it has not appeared outside the problem domain along with all of its immediate neighbors during the same time step. This can be prevented by placing an upper bound on the distance any boundary point moves during a single time step, expressed as constraint on the body CFL number $C_b = \max_s \norm{\vb{v}_b(s)} \Delta t / h$ where $\vb{v}_b(s)$ is the velocity of the moving boundary at surface coordinate $s$. Geometric arguments indicate that when $\Omega(t)$ is the region exterior to a convex body, $C_b < \sqrt{1/2}$ is sufficient. When $\Omega(t)$ is exterior to a nonconvex region, the body CFL constraint depends on the curvature of the boundary. Letting $\kappa$ be the largest negative (concave) curvature encountered on the boundary, a body CFL constraint of
\begin{equation}
    C_b < \sqrt{\frac{1}{2}} - \qty(\frac{1}{\abs{\kappa h}} - \sqrt{\frac{1}{\abs{\kappa h}^2} - \frac{1}{2}}) = \frac{1}{\sqrt{2}} - \frac{\abs{\kappa h}}{4} - \frac{\abs{\kappa h}^3}{32} + \order{\abs{\kappa h}^5}
\end{equation}
will guarantee that each point has a valid time history. This bound is obtained by taking $\Omega(t)$ to be a circle of radius $1/\abs{\kappa}$, and is less strict for well-resolved geometries with small $\kappa h$. }

The convergence behavior of \ReviewerOne{the update in Eq.~(\ref{eq:lsrk_moving})} is slightly more complex than the equivalent LSRK scheme applied to a method-of-lines discretization. Based on the numerical experiments described in section \ref{subsection:IIM_moving_1D_convergence} we observe that applying an $N$-th order LSRK scheme to an $M$-th IIM spatial discretization with moving boundaries yields an error of magnitude
\begin{equation}
    \norm{u - u_e} \sim \order{\Delta t^N} + \order{h^M} + \order{\Delta t h^{\tilde{M}}}.
\end{equation}
\ReviewerThree{Unlike a method of lines discretization, this treatment of moving bodies couples accuracy of the spatial and temporal discretizations, in that it contains a leading-order error term that cannot be additively separated into spatial and temporal contributions. Throughout this work we treat the coupled term primarily as a first-order temporal error with a prefactor proportional to $h^{\tilde{M}}$, with the} exponent $\tilde{M}$ determined by the order of spatial discretization and the order of the extension operators $E_f[\cdot]$ and $E_u[\cdot]$. We show below that in practice, when the extension operators are of order $M$ or higher and the \ReviewerThree{time step $\Delta t$} is limited by a \ReviewerThree{body CFL} criterion, \ReviewerThree{this} moving boundary error term is dominated by the spatial error term \ReviewerThree{and does not affect the overall convergence of the method}.

\subsection{1D Moving boundary temporal convergence} \label{subsection:IIM_moving_1D_convergence}
To illustrate the convergence behavior, consider an initial boundary value problem for the advection equation
\begin{equation}
        \pdv{u}{t} = c \pdv{u}{x} \qq{on} \Omega(t) = [0, \,1] \setminus [x_\alpha(t), x_\beta(t)], \\
\end{equation}
with solid boundaries translating at constant speed $v_b$ so that $x_\alpha(t) = x_{\alpha, 0} + v_b t$ and $x_\beta(t) = x_{\beta, 0} + v_b t$. We take the wave speed $c$ to be positive and greater than the boundary velocity $v_b$, so that an inflow boundary condition is required at $x_\beta(t)$ while $x_\alpha(t)$ acts as an outflow. The domain is periodic with $u(0, t) = u(1, t)$, while the initial condition $u(x, 0) = \sin(2\pi kx)$ and time-dependent inflow boundary condition $u(x_\beta(t), t) = \sin(2\pi k(x_\beta(t) - ct))$ are set to match the periodic exact solution $u_e(x, t) = \sin(2\pi k(x - ct))$. The spatial domain is discretized with $N_x$ uniformly spaced points so that $x_i = i/N_x$ for $i = 0$ to $N_x - 1$, and the time integration runs from $t = 0$ to $t = T$ with $N_t$ time steps of size $\Delta t = T / N_t$. The spatial derivative is discretized with the third order upwind finite difference scheme
\begin{equation}\label{eq:upwind_first_deriv}
    \qty(\pdv{u}{x})_i = \frac{u_{i-2} - 6u_{i-1} + 3 u_i + 2u_{i+1}}{6h} + \ReviewerThree{\order{h^3}}.
\end{equation}
\ReviewerTwo{At the moving boundaries this finite difference scheme is applied to an extended field constructed using the IIM methodology described in section \ref{subsection:IIM_fixed} and illustrated in Figure \ref{fig:extension_diagrams}. At a given time $t$, let $i_{\alpha}$ be the index of the grid point immediately left of the outflow boundary $x_{\alpha}(t)$, and let $i_{\beta}$ be the index of the grid point immediately right of the inflow boundary $x_{\beta}(t)$. The field $u(x, t)$ is extended to the grid point  $x_{i_{\alpha} + 1}$ using a fourth order IIM extension without boundary condition, and to the grid points $\{x_{i_{\beta} - 2},\, x_{i_{\beta} - 1}\}$ using a fourth order IIM extension that incorporates the prescribed inflow boundary condition at $x_{\beta}(t)$.} 
\ReviewerTwoStrike{After these extensions the finite difference stencil (\ref{eq:upwind_first_deriv}) can be applied to all points in the domain while retaining full third order accuracy and without affecting the observed stability of the method.} \ReviewerTwo{For simulations with stationary boundaries, this 1D advection discretization is shown to be third order accurate and stable under the CFL constraint $c \Delta x/\Delta t < 1.07$  in \cite{gabbard2023high}.} To integrate the semi-discrete system in time, we use low storage Runge-Kutta methods of order two (Heun's method) and three (\citet{Williamson1980}). The extension operator $E_f[\cdot]$ uses a third order IIM extension without boundary condition at both boundaries, while the extension operator $E_u[\cdot]$ uses a third order IIM extension with a boundary condition at $x_\beta(t)$ and without a boundary condition at $x_\alpha(t)$. For the numerical experiments shown here we take $c = 1$, $v_b = 0.25$, $k = 1$, $T = 0.7$, $x_{0,\alpha} = 0.261$, and $x_{0,\beta} = 0.447$. 

\begin{figure}[htb!]
    \centering
    \begin{subfigure}{0.46\linewidth}
        \centering
        \resizebox{\textwidth}{!}{\input{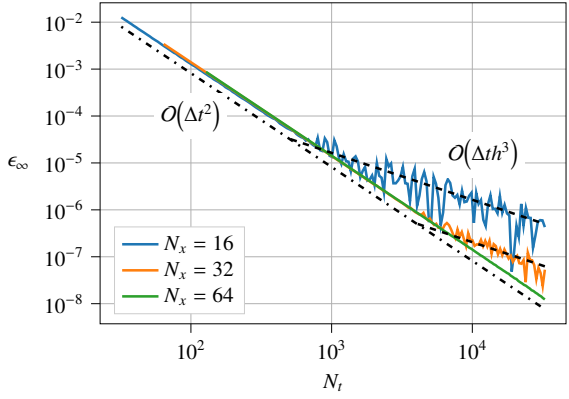}}
        \caption{Advection with low storage RK2.}
        \label{fig:moving_adv_error_rk2}
    \end{subfigure}
    \hspace{0.03\textwidth}
    \begin{subfigure}{0.46\linewidth}
        \centering
        \resizebox{\textwidth}{!}{\input{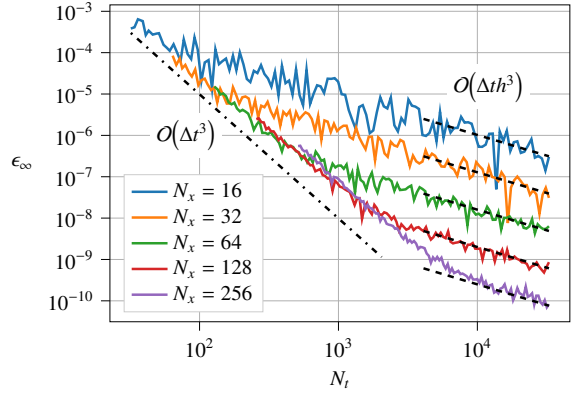}}
        \caption{Advection with low storage RK3.}
        \label{fig:moving_adv_error_rk3}
    \end{subfigure}
    \caption{Temporal convergence of the advection equation with moving boundaries at a fixed spatial resolution with an RK scheme of order two (a) or three (b). For each data point, the $L_\infty$ error is measured relative to a simulation with the same spatial resolution and a much finer temporal resolution ($N_t \approx 1.3 \times 10^5$). For a third order spatial discretization and $N$-th order RK scheme, the temporal error consists of an $\mathcal{O}(\Delta t^N)$ term which dominates at low temporal resolutions and an $\mathcal{O}(\Delta t h^3)$ that dominates at higher temporal resolutions. }
\end{figure}

To evaluate the temporal convergence of the moving boundary algorithm, the number of grid points $N_x$ is fixed while the number of time steps $N_t$ is varied across two orders of magnitude. For each simulation, the $L_\infty$ error in the solution $u(x, T)$ is evaluated relative to a solution generated with $N_{t, ref}$ time steps for $N_{t, ref}$ much larger than $N_t$. This eliminates any purely spatial errors from the comparison.
Figure \ref{fig:moving_adv_error_rk2} and \ref{fig:moving_adv_error_rk3} illustrate the resulting temporal convergence for low storage RK2 and RK3 integrators respectively; in both cases the convergence is initially $\mathcal{O}(\Delta t^N)$ for $N = 2, \, 3$ before dropping to $\mathcal{O}(\Delta t)$ as the step size increases. If the spatial resolution $N_x$ is varied, the magnitude of the initial $\mathcal{O}(\Delta t^N)$ error is unchanged, indicating that this is the purely temporal error of the RK scheme. The prefactor in the $\mathcal{O}(\Delta t)$ region varies with $h^3$, indicating a mixed error term of magnitude $\mathcal{O}(\Delta t h^3)$ due to the moving boundary treatment.

\begin{figure}[htb!]
    \centering
    \begin{subfigure}{0.46\linewidth}
        \centering
        \resizebox{\textwidth}{!}{\input{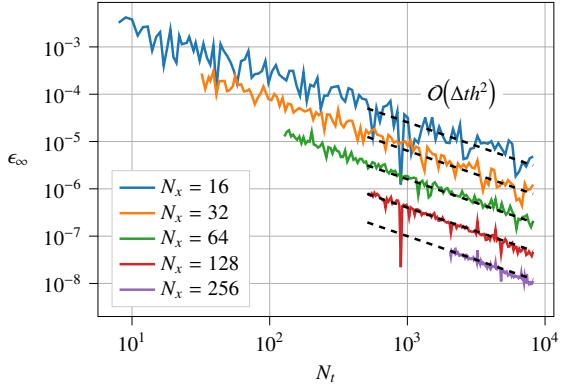}}
        \caption{Diffusion with low storage RK2.}
        \label{fig:moving_diff_error_22}
    \end{subfigure}
    \hspace{0.03\textwidth}
    \begin{subfigure}{0.46\linewidth}
        \centering
        \resizebox{\textwidth}{!}{\input{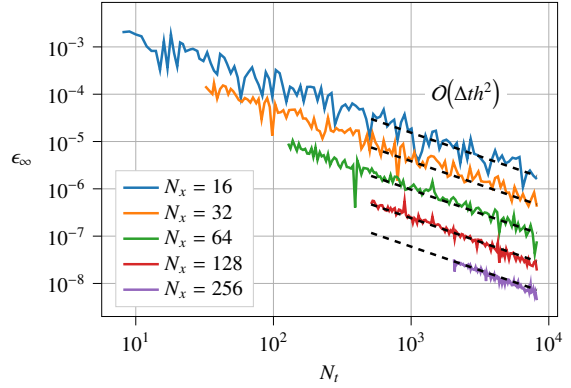}}
        \caption{Diffusion with low storage RK3.}
        \label{fig:moving_diff_error_44}
    \end{subfigure}
    \caption{Temporal convergence of the diffusion equation with moving boundaries at a fixed spatial resolution with an RK scheme of order two (a) or three (b). In both cases the error is dominated by an $\mathcal{O}(\Delta t h^2)$ moving boundary error term with a prefactor that is roughly the same for both integrators. }
\end{figure}

To further probe this behavior, the experiment above is repeated with an initial boundary value problem for the diffusion equation
\begin{equation}
    \pdv{u}{t} = \nu \pdv[2]{u}{x} \qq{on} \Omega(t)
\end{equation}
on the same periodic 1D domain with moving boundaries. The initial condition and Dirichlet boundary conditions at $x_\alpha(t)$ and $x_\beta(t)$ are set to match the exact solution $u_e(x, t) = \mathrm{exp}(-\nu k^2 t) \sin(2\pi kx)$. The spatial operator is discretized with a second order centered finite difference stencil (\ref{eq:centered_example_stencil}), and at both Dirichlet boundaries the field $u(x,t)$ is extended by one point using a fourth order IIM extension with boundary condition before the finite difference stencil is applied. We take $\nu = 0.01$ along with the parameter values used for the advection case. The temporal convergence of the scheme is evaluated as above with RK2 and low-storage RK3 time integration, using IIM extension operators $E_f[\cdot]$ of order two without boundary condition and IIM extension operators $E_u[\cdot]$ of order two with boundary condition. The convergence results for RK2 (Figure \ref{fig:moving_diff_error_22}) and low-storage RK3 (Figure \ref{fig:moving_diff_error_44}) both indicate a moving boundary error term of magnitude $\mathcal{O}(\Delta t h^2)$ which dominates the purely temporal error for all tested resolutions.

Generally, the observed convergence order for these numerical experiments can vary with the character of the PDE, the order of the free-space stencil, the order of the IIM boundary treatment, the order of the extension operators $E_f[\cdot]$ and $E_u[\cdot]$, and the order of the time integration scheme. We leave an in-depth investigation for future work, but note here that in practice, if the extension operators $E_f[\cdot]$ and $E_u[\cdot]$ are of sufficiently high order and the time step $\Delta t$ is limited by a CFL criterion, the moving boundary error term has magnitude $\mathcal{O}(h^{\tilde{M} + 1})$ and is dominated by the $\mathcal{O}(h^M)$ spatial error. This is corroborated by the numerical convergence tests shown in sections~\ref{section:results_oneWay} and \ref{section:results_twoWay}.

\subsection{2D Moving boundary temporal convergence}
\begin{figure}
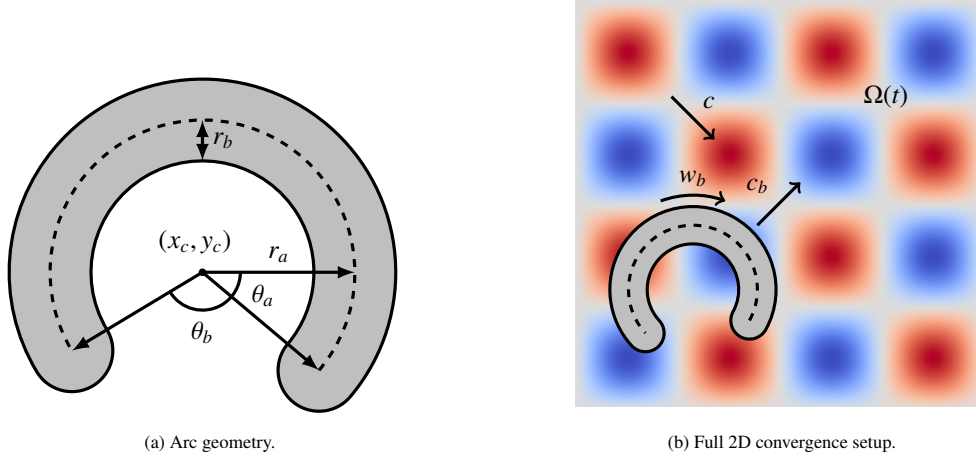

    \centering
    \begin{subfigure}{0.35\textwidth}
        \centering
        \resizebox{\textwidth}{!}{\input{tikzfigures/adv_diff_convergence_2d_geometry}}
        \caption{Arc geometry.}    
        \label{fig:adv_diff_2d_geometry}
    \end{subfigure}
    \hspace{0.1\textwidth}
    \begin{subfigure}{0.35\textwidth}
        \centering
        \resizebox{\textwidth}{!}{\input{tikzfigures/moving_setup_2d}}
        \caption{Full 2D convergence setup.}    
            \label{fig:adv_diff_2d_field}
    \end{subfigure}
    \caption{(a) Thickened arc geometry used for the 2D advection diffusion test case. For all results the shape parameters are held constant with $r_a = 0.1701$, $r_b = 0.0535$, and $\theta_b = 2.4$. The initial position is defined by the coordinates $x_0 = 0.287$, $y_0 = 0.289$ and the angular coordinate $\theta_a = 0.5$. Throughout the simulation the body translates with constant velocity $\vb{c}_b = [c_x, \, c_y]$ and rotates with constant angular velocity $w_b$, which are varied from test case to test case as described in the text. (b) Setup for the 2D moving convergence tests. \ReviewerThree{The moving domain $\Omega(t)$ is defined by the body velocity $\vb{c}_b$ and body angular velocity $w_b$. The flow velocity $\vb{c}$ for the through flow case is also depicted.}}
    \label{fig:adv_diff_setup}
\end{figure}

To test the convergence of the moving IIM discretization in 2D, we consider the advection diffusion equation
\begin{equation}
    \pdv{u}{t} + \nabla \cdot (\vb{c}  u) = \nu \nabla^2 u
\end{equation}
with constant velocity $\vb{c} = [c_x, \, c_y]$ and diffusivity $\nu$, posed on the periodic unit square $\Omega = [0, 1] \times[0, 1]$. The domain contains a single immersed body in the shape of a thickened arc, as shown in Figure \ref{fig:adv_diff_2d_geometry}, which provides an interface with both locally convex and locally concave regions. The arc translates with constant velocity $\vb{c}_b = [1, 1]$ and rotates about its center point with constant angular velocity $w_b = 2$. The manufactured solution 
\begin{equation}
    g(x, y, t) = \exp(-\nu(k_x^2 + k_y^2)t) \sin(k_x (x - c_x t)) \sin(k_y (y - c_y t))
\end{equation}
with $k_x = k_y = 4\pi$ is prescribed as both an initial condition and as a Dirichlet boundary condition on the immersed body, as shown in Figure \ref{fig:adv_diff_2d_field}. The PDE is discretized with the conservative finite-difference IIM developed in \cite{Gabbard2021}, which uses third order upwind advective fluxes and second order centered diffusive fluxes. Each test case is integrated in time from $t = 0$ to $t = 0.3$ with a third order low-storage RK scheme and a fixed time step, and the $L_\infty$ norm of the error at the final time is recorded. For each simulation the time step is set to be $75\%$ of the maximum linearly stable time step for the free space finite difference scheme, as determined by the method in \cite{Gabbard2021}. At moving boundaries the time integration is performed with extension operators $E_f[\cdot]$ of order two without boundary condition and extension operators $E_u[\cdot]$ of order three with boundary condition.

\begin{figure}[htb]
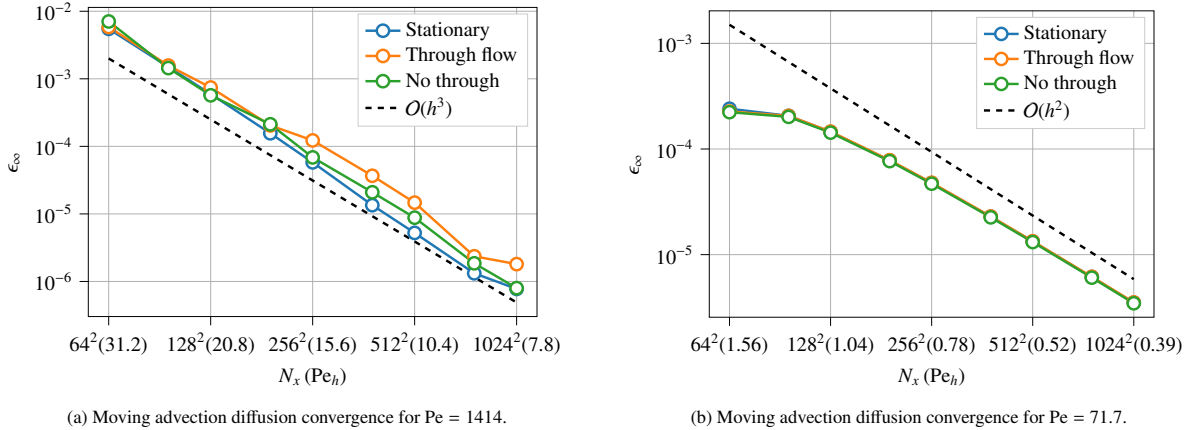

    \centering
    \begin{subfigure}{0.46\linewidth}
        \centering
        \resizebox{\textwidth}{!}{\input{tikzfigures/adv_diff_convergence_2d_high_pe}}
        \caption{Moving advection diffusion convergence for $\mathrm{Pe} = 1414$.}
        \label{fig:adv_diff_moving_high_pe}
    \end{subfigure}
    \hspace{0.03\textwidth}
    \begin{subfigure}{0.46\linewidth}
        \centering
        \resizebox{\textwidth}{!}{\input{tikzfigures/adv_diff_convergence_2d_low_pe}}
        \caption{Moving advection diffusion convergence for $\mathrm{Pe} = 71.7$.}
        \label{fig:adv_diff_moving_low_pe}
    \end{subfigure}
    \caption{Spatial convergence of a 2D IIM discretization of the advection-diffusion equation with stationary boundaries, moving through-flow boundaries, and moving no through-flow boundaries. The resulting error is third order when dominated the advection term and second order when dominated by the diffusion term, with the crossover occurring at cell Peclet number $\mathrm{Pe}_h \approx 1$. All three boundary types lead to roughly the same errors, indicating that the treatment of moving boundaries is not a dominant source of error. }
    \label{fig:adv_diff_2d_convergence}
\end{figure}

The qualitative nature of the solution depends on the Peclet number $\mathrm{Pe}_L = \norm{u}_2 L / \nu$, for which we use the domain length $L = 1$ as a reference scale. We consider here both advection-dominant cases with $\mathrm{Pe}_L = 1414$ and more diffusive cases with $\mathrm{Pe}_L = 71.7$. We also consider three different combinations of flow and body motion: a stationary reference case with $w_b = 0$, $\vb{c}_b = [0, 0]$ and $\vb{c} = [1, 1]$; a no through-flow case with $w_b = 0$ and $\vb{c} = \vb{c}_b = [1, 1]$; and a through flow case with $w_b = 2$, $\vb{c}_b = [1, 1]$, and $\vb{c} = [1, -1]$. Figures \ref{fig:adv_diff_moving_high_pe} and \ref{fig:adv_diff_moving_low_pe} plot the solution error as a function of the spatial resolution for the high and low $\mathrm{Pe}$ cases respectively. For all cases the convergence order depends on the cell Peclet number $\mathrm{Pe_h} = \norm{u}_1 h / \nu$ and mimics that of the free-space finite difference scheme: for $\mathrm{Pe}_h \gg 1$ the error is dominated by the third order advection term, and for $\mathrm{Pe}_h \ll 1$ the error is dominated by the second order diffusive term. There is little difference in the error between the stationary, through-flow, and no through-flow cases, demonstrating that the error introduced by moving boundaries is dominated by the existing spatial discretization error across a wide range of Peclet numbers and resolutions.

\section{One-way coupled vorticity-velocity based Navier-Stokes simulations}\label{section:algorithm}

\ReviewerTwo{Throughout this work we discretize the Navier-Stokes equations in vorticity-velocity form using a second-order IIM discretization originally developed for stationary bodies in \cite{Gabbard2021}. In this section we briefly review this algorithm while describing a novel extension to moving bodies with one-way coupling that makes use of the moving IIM approach developed in section \ref{section:IIM}.} \ReviewerTwo{Then we show the convergence and verification results of the one-way coupling method.}

\subsection{Navier-Stokes algorithm}\label{subsection:NS_fixed}
Taking the curl of the 2D incompressible velocity-pressure Navier-Stokes equations and invoking the continuity constraint $\nabla \cdot \vb{u} = 0$ yields the vorticity transport equation
\begin{equation}\label{eq:transport}
\pdv{\omega}{t} =  -\nabla \cdot (\vb{u}\omega - \nu \nabla \omega).
\end{equation}
This advection-diffusion equation requires a single boundary condition on all solid boundaries. Here we follow \citet{Gillis2019} and compute the boundary vorticity field directly from the velocity field, so that 
\begin{equation}
    \omega_b = \nabla \times \vb{u} \qq{on} \partial \Omega. 
\end{equation}
This is not the only possible choice, and we refer readers to \cite{Quartapelle1993} for a thorough discussion of numerical boundary conditions for the vorticity transport equation. Assuming that the volume of each immersed body is constant, \ReviewerThree{so that the boundary velocity $\vb{u}_b$ satisfies}
\begin{equation}\label{eq:constantvolume}
    \oint_{\partial B_k} \vb{u}_b \cdot \vb{n} \dd{s} = 0 \qq{for $1 \le k \le N_b$,}
\end{equation}
the velocity field $\vb{u}$ can be expressed in terms of a scalar stream function $\psi$ that obeys a Poisson equation with circulation constraints:
\begin{equation}
\begin{aligned}\label{eq:reconstruction}
    \vb{u} &= \nabla \times \psi \qq{on} \Omega, \\[0.5ex]
     -\nabla^2 \psi &= \omega  \qq{on} \Omega, \\[0.5ex]
     \psi &= \psi_{b,k} + \bar{\psi}_k \qq{on} \partial B_k,\\
     -\oint_{\ReviewerThree{C_k}} \partial_n \psi \dd{s} &= \Gamma_k \qq{for} 1 \le k \le N_b.
 \end{aligned}
\end{equation}
Here $\Gamma_k$ is the circulation \ReviewerThree{about a stationary closed contour $C_k$ encircling the $k$-th solid body}, $\bar{\psi}_k$ is an unknown constant associated with the $k$-th solid body, and $\psi_{b\ReviewerThree{,k}}$ is any fixed boundary condition that satisfies $\partial_s \psi_{b\ReviewerThree{,k}} = \vb{u}_b \cdot \vu{n}$ on \ReviewerThree{$\partial B_k$}. Condition (\ref{eq:constantvolume}) ensures that it is always possible to construct a single-valued function $\psi_{b\ReviewerThree{,k}}$ \ReviewerThree{defined on each solid boundary by the integral}
\begin{equation}
    \psi_{b\ReviewerThree{,k}}(s) = \int_{\ReviewerThree{s_{0,k}}}^s \vb{u}_b(s) \cdot \vu{n} \dd{s} \ReviewerThree{\qq{for} s \in \partial B_k.}
\end{equation}
\ReviewerThree{Here each} integral is taken counterclockwise over \ReviewerThree{the boundary $\partial B_k$} beginning at an \ReviewerThree{arbitrary boundary point $s_{0,k}$}. The full velocity reconstruction problem is completed by an additional boundary condition on the far field or on the edge of the computational domain; appropriate conditions for periodic, unbounded, and interior domains are discussed in section 4.1 of \cite{Gabbard2021}. The system is closed by \ReviewerThree{applying Kelvin's circulation theorem to each of the closed contours $C_k$, yielding a relationship between the circulations $\Gamma_k$ and the flux of vorticity across the contour:}
\begin{equation}\label{eq:kelvin}
    \dv{\Gamma_k}{t} = -\oint_{\ReviewerThree{C_k}} (\vb{u}\omega - \nu \nabla \omega) \cdot \vu{n} \dd{s}.
\end{equation}
This constraint is obtained by integrating the tangential component of the velocity-pressure Navier-Stokes equations around \ReviewerThree{$C_k$}, and represents a separate constraint that cannot be derived directly from the vorticity transport equation.

\ReviewerTwo{The discretization of Equations (\ref{eq:transport}) through (\ref{eq:kelvin}) is taken directly from the second-order IIM developed for flows with stationary bodies in \cite{Gabbard2021}}. As in that work, the simulation state is captured by the discretized vorticity field and a set of circulations $\{\Gamma_k\}$, one for each immersed body, which are defined as the circulation of the velocity field around a bounding box \ReviewerThree{$C_k$} (here termed the ``circulation box") associated with each body. \ReviewerTwo{Both the vorticity field and the circulations $\{\Gamma_k\}$ are evolved in time with a second or third order Runge-Kutta time integrator. The main difference with \cite{Gabbard2021} is that during time integration, we apply the moving boundary treatment outlined in section \ref{subsection:IIM_moving} at solid boundaries. Here the operators $E_{u}[\cdot]$ applied to the vorticity field use a third-order IIM extension that incorporates the vorticity boundary condition $\omega_b$, while the operators $E_{f}[\cdot]$ applied to the time derivative $\partial_t \omega$ use a second order IIM extension without boundary condition. }.

\ReviewerTwoStrike{The spatial discretization begins by solving for the stream-function $\psi$ and unknown constants $\bar{\psi}_k$ using the second-order IIM Poisson solver outlined in  section 4.2 of \cite{Gabbard2021}. The velocity field is calculated from $\psi$ using a second-order centered finite difference scheme, with a third-order IIM extension that incorporates the boundary condition on $\psi$ given in Eq.~(\ref{eq:reconstruction}). The wall vorticity  $\omega_b$ at the IIM control points is calculated by evaluating the curl of the IIM-extended velocity field on the solid boundaries, where the extension incorporates the no-slip condition as a boundary condition; see  section 5.3 of \cite{Gabbard2021}. The right hand side of the vorticity transport equation (\ref{eq:transport}) is discretized using the second-order conservative finite difference scheme and the IIM boundary treatment for advection-diffusion equations discussed in section 3.2 of \cite{Gabbard2021}. Finally, the right hand side of (\ref{eq:kelvin}) is calculated by summing the vorticity flux from the conservative difference scheme over the circulation box associated with each immersed body. Once these calculations are complete, both the vorticity fields and the circulations are updated using the same Runge-Kutta time integrator.}

\subsection{Force calculation}\label{subsection:Force}
The use of the vorticity-velocity formulation in this work means that the surface pressure field is not directly available for force and torque calculations. \ReviewerTwo{Instead, we use a control volume approach\footnote{Since we are in 2D, this could be more appropriately denoted a ‘control area’} that does not rely on pressure to obtain body forces and moments. This method was first introduced in \cite{Noca1997}, then widely used in the force calculation of the immersed boundary method~\cite{shen_calculation_2009, bergmann_bioinspired_2016, nangia_moving_2017}. This was further applied to the immersed interface method in  \cite{Gabbard2021}. In this section, we briefly review the force calculation of \cite{Gabbard2021} in preparation for our two-way coupling method in section~\ref{section:NS_twoWay}.} \ReviewerTwoStrike{Instead, as in \cite{Gabbard2021}, we use a control volume approach that does not rely on pressure \cite{Noca1997} to obtain body forces and torques.} Specifically, we create a grid-aligned control volume around each immersed body as shown in Figure~\ref{fig:forcebox}. We use the ‘momentum 4’ approach from \citep{Noca1997} to find expressions for the body force given integral quantities over the area and boundaries of this control volume:

\begin{figure}[htb]
    \centering
    \includegraphics[width=0.25\textwidth]{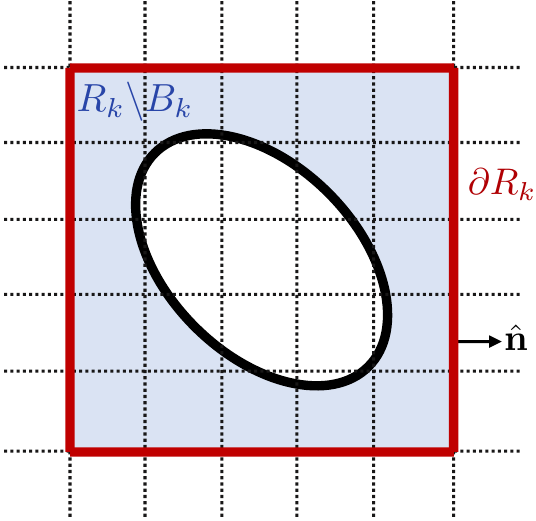}
    \caption[]{The configuration of a control volume for computing forces and torques on immersed bodies. Black dotted lines
    (\tikz[baseline] \draw[black, dotted, very thick] (0pt, 2pt) -- (7pt, 2pt);)
    are grid lines, the black solid line 
    (\tikz[baseline] \draw[black, very thick] (0pt, 2pt) -- (7pt, 2pt);)
    is the boundary of the body, the red solid line 
    (\tikz[baseline] \draw[red, very thick] (0pt, 2pt) -- (7pt, 2pt);)
    is the boundary of the force box $\partial R_k$, the blue region represents $R_k\backslash B_k$ for the area integral, and the black arrow shows the direction of the normal unit vector $\vu{n}$.}
    \label{fig:forcebox}
\end{figure}

\begin{equation}
    \vb{F}_f = -\dv{}{t}\int_{R_k\backslash B_k} \vb{u} \dd{A} - \dv{}{t}\oint_{\partial R_k} \vb{x} \times (\vu{n} \times \vb{u}) \dd{s} + \oint_{\partial R_k} \vu{n} \cdot \bm{\gamma} \dd{s},\label{eq:fluidForce}
\end{equation}
where $\vu{n}$ is the normal unit vector, $R_k$ is the region covered by the force box, which is identical to the circulation box in all our simulations, $B_k$ is the region covered by the body, and $\partial R_k$ represents the boundary of the force box. Here we only consider 2D domains, so the integrals over $R_k\backslash B_k$ and $\partial R_k$ are area and boundary integrals respectively. The quantity $\bm{\gamma}$ is a tensor with miscellaneous terms measured on the boundary of the force box:

\begin{equation}
    \bm{\gamma} = \frac{1}{2}|\vb{u}|^2\mathcal{I} - \vb{u}\vb{u}-\vb{u}(\vb{x}\times \omega \vu{k}) + \vb{x} \cdot(\nabla \cdot \vb{T})\mathcal{I} - \vb{x}(\nabla \cdot \vb{T}) + \vb{T},
\end{equation}
where $\mathcal{I}$ is the identity tensor, $\vb{T} = \nu(\nabla \vb{u} + \nabla \vb{u}^T)$ is the viscous stress tensor, and $\vu{s} = \vu{k} \times \vu{n}$ is the tangential unit vector. Similarly, the fluid moment on the obstacle about the origin $(0,0)$ is \cite{Gabbard2021},

\begin{equation}
    M_{f,0} = -\dv{}{t}\int_{R_k\backslash B_k} \vb{x} \times \vb{u} \dd{A} + \dv{}{t}\oint_{\partial R_k} \frac{\abs{\vb{x}}^2}{2} \vu{n} \times  \vb{u} \dd{s}  + \oint_{\partial R_k} \bm{\lambda}\dd{s},
    \label{eq:fluidMoment_origin}
\end{equation}
where the quantity $\bm{\lambda}$ collects miscellaneous terms on the force box boundary:

\begin{equation}
    \bm{\lambda} = \frac{1}{2}|\vb{u}|^2(\vb{x} \times \vu{n}) - (\vb{x} \times \vb{u})(\vb{u} \cdot \vu{n}) - \frac{|\vb{x}|^2}{2}\vu{n} \times (\vb{u} \times \omega \vu{k}) + \frac{|\vb{x}|^2}{2}(\nabla \cdot \vb{T})\times \vu{n} + \vb{x} \times (\vb{T} \cdot \vu{n}).
\end{equation}
The moment about the origin $M_{f,0}$ can be be transferred to a moment around the center of rotation of the body $\vb{x}_c$ by replacing $\vb{x}$ with $\vb{x} - \vb{x_c}$ in the equations above. We note that none of these expressions require boundary integrals over the body surface, which simplifies their implementation in our specific context compared to some of the other approaches discussed in \cite{Noca1997}, or used in \cite{Eldredge2007}. 

\ReviewerTwoStrike{In all our simulations,} \ReviewerTwo{As in \cite{Gabbard2021},} the line integrals are discretized with the trapezoidal rule. The area integral excludes the obstacle region inside the control volume and is discretized by combining a polynomial extrapolation with the second order level-set integration method from Towers \cite{Towers2009}. Time derivatives of these integrals are computed in post-processing using central differences. 

For simulation cases with a single obstacle, the control volume can cover the whole domain. With multiple bodies in the same domain, we choose the size of the control volume around each body to be relatively large, e.g.\ for a cylinder at least one diameter away from the perimeter. Larger control volumes in practice significantly reduce spurious noise in the force measurements, especially when the flow field within the region exhibits strong dynamic changes. \ReviewerTwoStrike{For moving bodies,} \ReviewerTwo{Specific to the moving bodies considered in this work,} the control volumes are regularly (every 10-20 timesteps) reset to match the updated location of the body. We record the changes and, in post-processing, perform one-sided differences to compute time derivatives immediately before and after the resets.

\subsection{Pressure distribution}\label{subsection:Pressure}
The control volume approach provides the overall force and torque coefficients, but does not provide insight into the distribution of surface tractions $\vb{t} = -p \vu{n} + \tau \vu{s}$, with $p$ surface pressure and $\tau$ the surface shear stress. The shear stress distribution can be directly evaluated from the wall vorticity : $\tau(s) = \nu (\omega_b(s) - 2 \Omega_b)$, where $\Omega_b$ is the angular velocity of the body. To compute the pressure distribution, we can evaluate the Navier-Stokes equations at the boundary:
\begin{equation}
    \ReviewerTwo{\frac{\partial \vb{u}}{\partial t} + (\vb{u} \cdot \nabla) \vb{u} = -\nabla p + \nu \nabla^2 \vb{u}},
    \label{eq:NS}
\end{equation}
 \ReviewerTwo{where $\nu \nabla^2 \vb{u}$ is equivalent to $-\nu \nabla \times \omega$ since $\omega = \nabla \times \vb{u}$ and $\nabla \cdot \vb{u} = 0$. In a frame moving with the boundary, equation~\eqref{eq:NS} can be written as:}
 \begin{equation}
     \ReviewerTwo{\dv{\vb{u}}{t} = -\nabla p - \nu \nabla \times \omega,}
 \end{equation} 
 \ReviewerTwo{where $\dd / \dd t$ is the Lagrangian time derivative.} If a no-slip boundary condition is enforced on the body, then the fluid acceleration is identical to the body acceleration $\vb{a}_b(s)$. The tangential component of this equation can then be written as:
\begin{equation}
    \frac{\partial p}{\partial s} = \nu \frac{\partial \omega}{\partial n} - \vb{a}_b \cdot \hat{\vb{s}},
\end{equation}
which relates the pressure gradient to the normal vorticity flux. From the vorticity flux we can find $p(s)$ by choosing an arbitrary point at $s_0$ on the boundary and integrating around the body surface:
\begin{equation}
    p(s) = p(s_0) + \int_{s_0}^s\left(  \nu \frac{\partial \omega}{\partial n} - \vb{a}_b \cdot \hat{\vb{s}} \right),
\end{equation}
where the normal vorticity flux is numerically obtained as sketched in \cite{Gabbard2021}, section 5.4. Note that this finds the pressure distribution only up to an arbitrary, potentially time-varying constant.

\subsection{One-way coupling results}\label{section:results_oneWay}

Here we discuss convergence and verification results for the presented Navier-Stokes solver with one-way coupled bodies, i.e.\ bodies with prescribed kinematics. All results are created with a second-order Runge-Kutta time integration scheme using a timestep that is 70\% of the maximum possible stable step given the explicit treatment of the viscous and advective terms \cite{Gabbard2021}; in practice this leads to a CFL number around 0.5. Below we first show the convergence of our approach using a test case of an impulsively moving cylinder. Then we demonstrate our algorithm on a pitching plate case and compare the results and convergence with a first-order penalization approach. Finally, we simulate a linearly oscillating cylinder to demonstrate a long-time simulation as well as the computed pressure distribution for accelerating obstacles.

\subsubsection{Convergence: impulsively started cylinder}\label{section:movingCylinder}
\begin{figure}
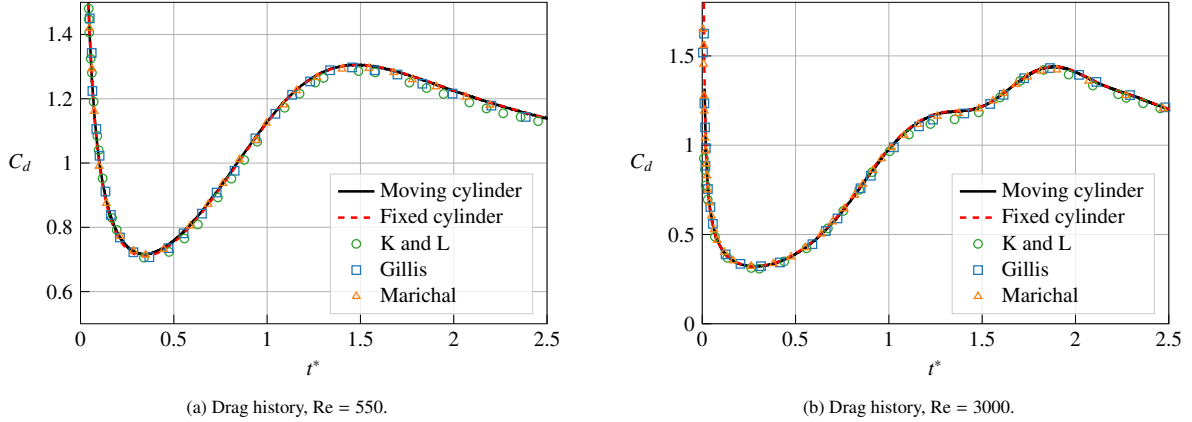

    \centering
    \begin{subfigure}{0.46\linewidth}
        \centering
        \resizebox{\textwidth}{!}{\input{tikzfigures/movCylinder_Re550}}
        \caption{Drag history, $\Re = 550$.}
        \label{fig:Movcylinder550}
    \end{subfigure}
    \hspace{0.03\textwidth}
    \begin{subfigure}{0.46\linewidth}
        \centering
        \resizebox{\textwidth}{!}{\input{tikzfigures/movCylinder_Re3000}}
        \caption{Drag history, $\Re = 3000$.}
        \label{fig:Movcylinder3000}
    \end{subfigure}
    \caption{Drag history for the impulsively moving cylinder at (a) $\Re = 550$ and (b) $\Re = 3000$. }
\end{figure}

The short-term forces on a stationary cylinder in an impulsively started rectilinear flow are frequently used to compare results between vorticity-velocity based approaches. Here we change the frame of reference to move the cylinder with an impulsively started constant rectilinear motion in an otherwise stationary fluid. Specifically, a cylinder of diameter $D$ and initial center $\vb{x}_c = (x_c, y_c)$ is placed in a stationary flow with kinematic viscosity $\nu$. At $t=0$, the cylinder begins moving with a constant speed $\vb{u}_b = (u_{b,x}, u_{b,y})$. The dynamics of the flow for $t>0$ depend only on the Reynolds number $\text{Re} = u_b D/\nu$, and the normalized time $t^* = u_bt/D$, where $t^*$ is in a time interval $[t_0^*, t_f^*]$. In order to first validate the one-way coupling method, here we consider an impulsively started cylinder with parameters in Table \ref{tab:MovCylinder}, where $\vb{U}_{\infty}$ is the velocity of fluid flow, $N^* = D/h$ represents inverse of the non-dimensional grid spacing. The drag coefficient, $C_d = 2F_x /(\rho_f u_b^2D)$, is compared with results of the impulsively started flow around a stationary cylinder from our previous work \cite{Gabbard2021}, Gillis \citep{GillisThesis}, Marichal \citep{MarichalThesis}, and Koumoutsakos and Leonard (K and L) \citep{PandL1995}.  We choose $N^*$ identical to our previous study of a stationary cylinder in \cite{Gabbard2021}, leading to $N^* = 204.8$ for $\text{Re} = 550$ and $N^* = 409.6$ for $\text{Re} = 3000$. Figures \ref{fig:Movcylinder550} and \ref{fig:Movcylinder3000} show that the results of the moving cylinder and stationary cylinder are in good agreement for both Reynolds numbers, and that our results also match well with the reference results.

\begin{table}[htb]
    \centering
    \begin{tabular}{ | c | c |} 
    \hline
    Element & Parameters  \\ 
    \hline\hline
    Grid & $\vb{x}_{ij} = (i/N_x, j/N_y)$ for $0 \leq i \leq N_x - 1$, $0 \leq j \leq N_y - 1$. $N_x = 2N_y$\\
    \hline
    Time & $t_0^* = 0.0$, $t_f^* = 3$\\
    \hline
    Obstacle & $\vb{x}_c = (3.75D, 1.2525D)$, $D = 0.4$\\
    \hline
    Motion & $\vb{u}_b = (-1.0, 0.0)$\\
    \hline
    Flow & $\vb{U_{\infty}} = (0.0, 0.0)$\\
    \hline
    \end{tabular}
    \caption{Simulation parameters for the impulsively moving cylinder case, where $N_x = N^*/D$ with $N^* = D/h$ is the number of grid points along the diameter.}
    \label{tab:MovCylinder}
\end{table}

\begin{figure}
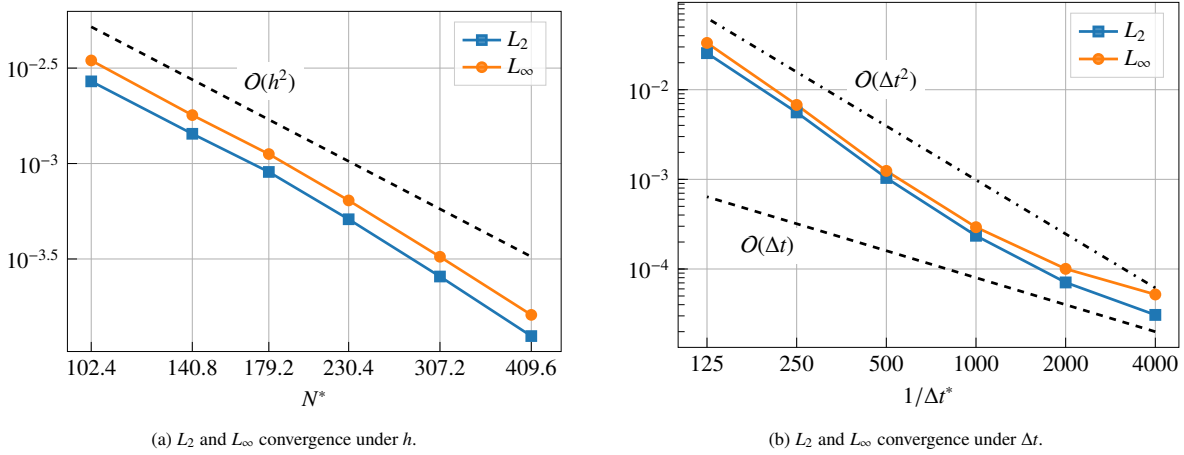

    \centering
    \begin{subfigure}{0.46\linewidth}
        \centering
        \resizebox{\textwidth}{!}{\input{tikzfigures/movCylinder_spatial_rk2}}
        \caption{$L_2$ and $L_{\infty}$ convergence under $h$.}
        \label{fig:MovCylinder_spatial}
    \end{subfigure}
    \hspace{0.03\textwidth}
    \begin{subfigure}{0.46\linewidth}
        \centering
        \resizebox{\textwidth}{!}{\input{tikzfigures/movCylinder_temporal_rk2}}
        \caption{$L_2$ and $L_{\infty}$ convergence under $\Delta t$.}
        \label{fig:Movcylinder_temporal}
    \end{subfigure}
    \caption{Spatial and temporal convergence for the impulsively moving cylinder at $\Re = 550$. }
\end{figure}

We proceed by using this test case to demonstrate the convergence rate of the one-way coupling algorithm. Here we define the norms of error in $[t_0^*, t_f^*]$,

\begin{align}
\centering
     & L_2 = \sqrt{\frac{1}{t_f^* - t_0^*} \int_{t_0^*}^{t_f^*}(f(t) - f_{ref}(t))^2 \text{d} t},  \label{eq:timeL2}\\
     & L_{\infty} = \text{max}\abs{f(t) - f_{ref}(t)}, \label{eq:timeLinf}
\end{align}

where $f(t)$ is the quantity for evaluation, and $f_{ref}(t)$ is the reference data. We take $t_0^* = 1.0$, $t_f^* = 2.0$, and $f(t) = I_x^*(t)$, where $I_x$ is the x-direction component of $\vb{I}$ in Section \ref{section:NS_twoWay}, which is normalized as $I_x^* = I_x/(D^2u_b)$. We run our convergence test with constant CFL, so that $\Delta t \sim \mathcal{O}(h)$, and at dimensionless grid spacing $N^* = D/h$ ranging from $N^* = 102.4$ to $N^* = 409.6$. Results at resolution $N^* = 819.2$ are taken as a reference for the error computation. We set $\text{Re} = 550$, and keep all the other settings the same as Table \ref{tab:MovCylinder}. From the convergence plot in Figure \ref{fig:MovCylinder_spatial}, we can conclude that our one-way coupling method achieves second order spatial convergence, as expected.

To measure temporal convergence, we consider the same case as above but now fix the resolution at $N^* = 51.2$ and decrease the non-dimensional timestep $\Delta t^* = u_b \Delta t/D$ from $8.0\times 10^{-3}$ to $2.5\times 10^{-4}$. The solution at $\Delta t^* = 2.5\times 10^{-5}$ is taken as reference for computing the error in $I_x^*$. Figure \ref{fig:Movcylinder_temporal} shows that we achieve second-order temporal convergence for larger timesteps, reducing to first-order convergence at small timesteps. This is consistent with the results in section~\ref{subsection:IIM_moving_1D_convergence}: for large timesteps we observe the free-space temporal convergence error from the RK2 time integrator, at very small timesteps we observe an $\mathcal{O}(\Delta t h^2)$ mixed error term.

Overall, the convergence results for one-way coupled bodies imply that at practical timestep values, our approach to moving bodies has the same convergence behavior as our previous second-order IIM for stationary bodies \cite{Gabbard2021}.

\subsubsection{Pitching plate}
To demonstrate our algorithm on a test problem with more practical relevance, we reproduce one of the pitching plate cases discussed in \citep{EandW2010} in the context of leading-edge vortex dynamics. The body is a thin plate of chord length $c$ and thickness $0.023c$, with semicircular edges. The plate pitches around its leading edge $\vb{x}_c$ in an incompressible flow with Reynolds number $Re = U_{\infty}c/\nu = 1000$, where $U_{\infty}$ is the velocity of the incoming flow. The angle of attack, $\alpha$, varies dynamically according to the function

\begin{equation}\label{alphat}
    \alpha(t) = \alpha_0\frac{G(t)}{G_{\max}},
\end{equation}
where the maximum angle, $\alpha_0$, is fixed to 45 degrees. The function $G(t)$ describes a smooth pitch-up motion
\begin{equation}\label{G}
    G(t) = \ln\left[\frac{\cosh{(a_sU_{\infty}(t - t_1)/c)}}{\cosh{(a_sU_{\infty}(t - t_2)/c)}}\right] - a_sU_{\infty}(t_1 - t_2)/c, \quad \quad \to G_{\max} = 2a_s(t_2 - t_1),
\end{equation}
where the dimensionless parameter $a_s$ controls the speed of transitions between kinematic intervals; here $a_s = 11$ throughout. The times $t_1$ and $t_2$ control the time of transition during the pitching motion. Following \cite{EandW2010} we place the start of the pitch-up at time $t_1 = c/U_{\infty}$ and parametrize the end time $t_2 = t_1 + \alpha_0/\Dot{\alpha}_0$ with a non-dimensional pitch-rate defined by $K= \Dot{\alpha}_0c/(2U_{\infty})$.

\begin{figure}
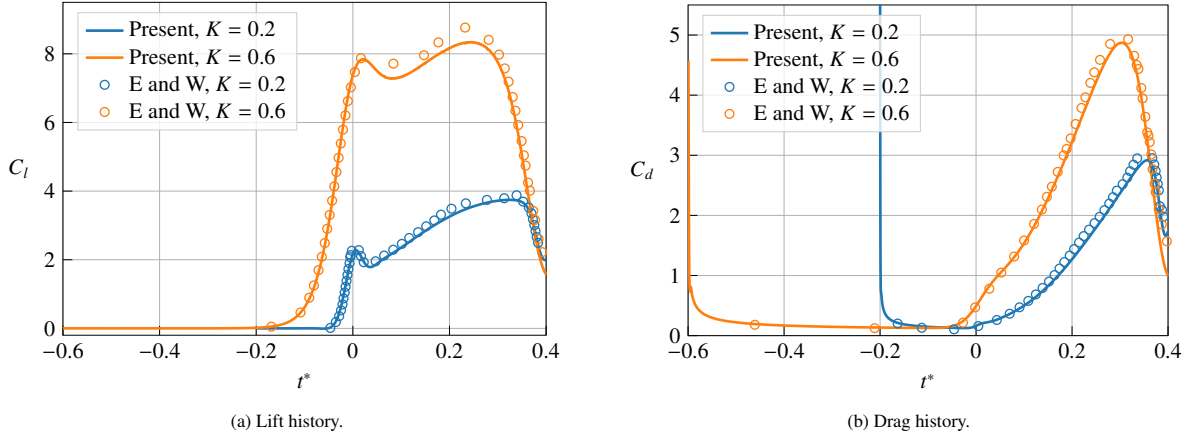

    \centering
    \begin{subfigure}{0.46\linewidth}
        \centering
        \resizebox{\textwidth}{!}{\input{tikzfigures/pitch_cl}}
        \caption{Lift history.}
    \end{subfigure}
    \hspace{0.03\textwidth}
    \begin{subfigure}{0.46\linewidth}
        \centering
        \resizebox{\textwidth}{!}{\input{tikzfigures/pitch_cd}}
        \caption{Drag history.}
    \end{subfigure}
    \caption{Lift and Drag history for the pitching plate at $K = 0.2$ (blue) and $K = 0.6$ (orange), where the non-dimensional start time of each simulation is $t^* = -K$. Present results are in solid lines, and results from Eldredge and Wang (E and W) \citep{EandW2010} are in circles. }
    \label{fig:pitchresults}
\end{figure}

We simulated cases with $K = 0.2$ and $K = 0.6$ using the parameters in Table \ref{tab:PitchingPlate}. The resulting lift and drag are normalized as $C_l = 2F_y  /(\rho_f U_{\infty}^2c)$ and $C_d = 2F_x /(\rho_fU_{\infty}^2c)$, and time is normalized as $t^* = K(tU_{\infty}/c - 1)$ so that the simulations start at $t^* = -K$. Figure~\ref{fig:pitchresults} shows a comparison between results of \citep{EandW2010} and our own results at $N^*=512$, demonstrating good agreement between the two results. 

\begin{table}[htb]
    \centering
    \begin{tabular}{ | c | c |} 
    \hline
    Element & Parameters  \\ 
    \hline\hline
    Grid & $\vb{x}_{ij} = (3i/N_x, 3j/N_y)$ for $0 \leq i \leq N_x - 1$, $0 \leq j \leq N_y - 1$. $N_x = N_y$\\
    \hline
    Time & $t_0 = 0.0$, $t_f = 2.0$\\
    \hline
    Obstacle & $\vb{x}_c = (0.51, 2.0)$, $c = 1.0$\\
    \hline
    Flow & $\vb{U_{\infty}} = (1.0, 0.0)$, $\nu = 1.0 \times 10^{-3}$\\
    \hline
    \end{tabular}
    \caption{Simulation parameters for the pitching plate case, where $N_x = 3N^*$ with $N^* = c/h$ is the number of grid points along the chord.}
    \label{tab:PitchingPlate}
\end{table}

To compare the effect of our second-order interface treatment with a first-order smooth immersed boundary treatment, we simulate the same pitching plate test case with a volume penalization method \citep{Angot:1999, Gazzola:2011a}. Specifically, we implemented the penalization algorithm presented in \cite{Bernier2019} using the same free-space spatial discretization operators as our IIM implementation. As reported in \citep{Gazzola:2011a, Bernier2019}, the resulting algorithm has convergence rates that are first-order in space and time in the infinity norm. 

\begin{figure}
    \centering
    \begin{subfigure}[t]{0.3\linewidth}
        \centering
        \resizebox{\textwidth}{!}{\input{tikzfigures/pitch_resolution_large}}
        \caption{Lift history for the pitching plate at $K = 0.6$ compared between the IIM method (IIM) and the penalization method (Penal) at various values of non-dimensional resolution $N^*$.}
    \end{subfigure}
    \hspace{0.02\textwidth}
    \begin{subfigure}[t]{0.3\linewidth}
        \centering
        \resizebox{\textwidth}{!}{\input{tikzfigures/pitch_resolution_cl}}
        \caption{Zoom-in results of panel (a) around the time of the second peak in the lift coefficient.}
    \end{subfigure}
    \hspace{0.02\textwidth}
    \begin{subfigure}[t]{0.32\linewidth}
        \centering
        \includegraphics[width=\textwidth]{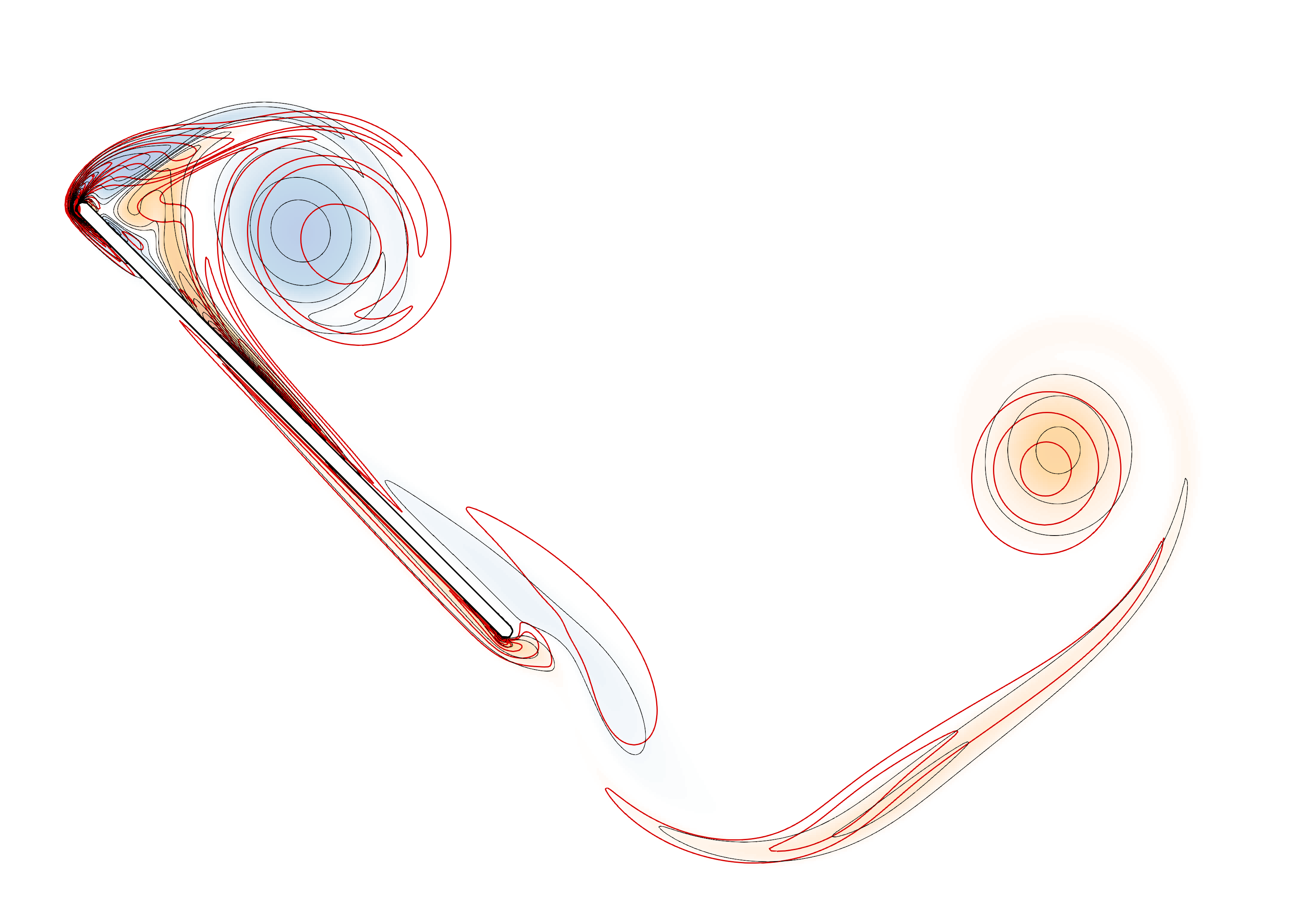}
        \caption{Contours of the vorticity field at \ReviewerOne{$t^* = 0.6$} for IIM (black) and the penalization method (red), both at $N^*=256$. The background colors represent the vorticity field of the IIM simulation at this time for reference. }
    \end{subfigure}    
    \caption{Comparison of IIM and penalization results for the pitching plate benchmark case proposed in \cite{EandW2010}.}
    \label{fig:pitchreso}
\end{figure}

We used both the penalization method and our current immersed interface method to run the pitching plate simulation with settings as in Table~\ref{tab:PitchingPlate}, at different spatial resolutions. For all penalization simulations, we used the same timestep criterion as for the IIM simulations, and set the non-dimensional penalization parameter $\lambda c / U_\infty = 10^5$ --- in-line with best practice values used in literature \citep{Gazzola:2011a, Rossinelli2015, Bernier2019}. Figure~\ref{fig:pitchreso} shows the computed lift coefficient evolution for both methods at different resolutions. We note that the IIM results are converged at $N^* = c/h = 256$, whereas the penalization results are still changing substantially even up to $N^* = 1024$, the highest resolution considered.

To quantify this behavior we define the non-dimensional vorticity error field as 
\begin{align}
\centering
e(\vb{x}_{ij}, t^*) = \frac{| \omega(\vb{x}_{ij}, t^*) - \omega_{ref}(\vb{x}_{ij}, t^*) |}{\omega_{ref,\max}(t^*)},
\end{align}
where $\omega_{ref}$ is the reference vorticity field. Since we do not have an exact solution, we use the IIM-computed vorticity field at resolution $N^* = 1024$ as the reference for both methods. This is motivated by the fact that if both methods represent convergent discretizations of the same continuous equations, and the IIM has a demonstrably higher convergence rate than the penalization approach, then the IIM field at $N^* = 1024$ represents the most accurate solution available. The errors are evaluated at time $t^* = 0.0$, and all spatial resolutions are chosen so that any grid point location at a lower resolution is also available in the reference field. Lastly, since the penalization method does not offer a sharp boundary treatment, the errors for all cases are computed only at grid points that are farther than $0.01c$ from the body surface \ReviewerOne{or outside of the penalized mollification region which spans a distance of $\sqrt{2} h$ outside of the body, whichever is largest}. 

\begin{figure}
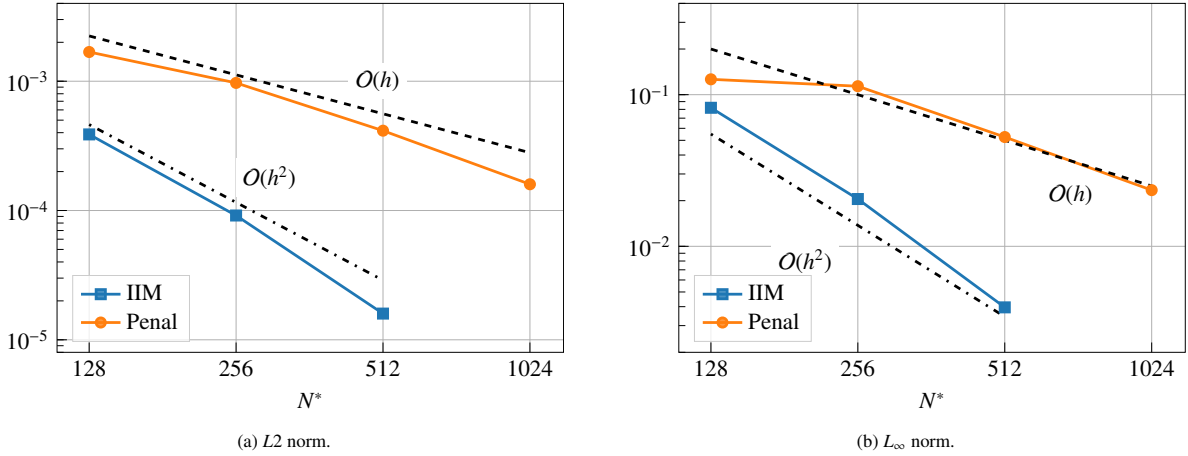

    \centering
    \begin{subfigure}{0.46\linewidth}
        \centering
        \resizebox{\textwidth}{!}{\input{tikzfigures/pitch_L2}}
        \caption{$L2$ norm.}
    \end{subfigure}
    \hspace{0.03\textwidth}
    \begin{subfigure}{0.46\linewidth}
        \centering
        \resizebox{\textwidth}{!}{\input{tikzfigures/pitch_Linf}}
        \caption{$L_{\infty}$ norm.}
    \end{subfigure}
    \caption{Vorticity field error convergence plot from IIM method (IIM) and penalization method (Penal) at $t^* = 0$.}
    \label{fig:pitchconverge}
\end{figure}

Figure \ref{fig:pitchconverge} plots the $L_2$ and $L_\infty$ norms of this error field at resolutions $N^* = 128$ to $N^* = 512$ for the IIM method, and $N^* = 128$ to $N^* = 1024$ for the penalization method. The results confirm the second-order convergence rate of IIM in both the $L_2$ and $L_\infty$ norm. The penalization method is first order in the $L_\infty$ norm, and somewhere between first and second order in the $L_2$ norm - consistent with previous findings \citep{Gazzola:2011a, Bernier2019} and theoretical estimates. Directly comparing the error values between the two approaches at these resolutions, we find that for the same accuracy the IIM approach requires about four to five times less grid points per dimension, or about 20 times less grid points in total.  This emphasizes the results shown in  Figure~\ref{fig:pitchreso}, namely that the second-order approach significantly improves the fidelity of simulations at practical grid resolutions. 

We refrain from directly comparing computational performance as our penalization implementation is not optimized. We do note that for both the penalization and the IIM approaches, the free-space Poisson solve takes the most amount of time per timestep. In the penalization algorithm, we perform two free-space Poisson solves per timestep; in IIM, we perform four. Together with other overheads related to the sharp-interface treatment, we would expect that for the same resolution grid our IIM solver would be roughly 3 times more expensive per timestep than the penalization implementation. The above convergence plot indicates that these expenses are more than offset by the gains in resolution requirements for a given error level, which motivates the use of even higher-order discretization approaches in the future.

\subsubsection{Pressure distribution: oscillating cylinder}
To examine the ability of our method to simulate long-time flow evolutions, we consider the diagnostics of a cylinder of diameter $D$ oscillating transversely in a flow with uniform velocity $U_{\infty}$ in the $x$-direction. The cylinder's $y$-direction displacement is $\Delta y(t) = -A \cos{(2\pi f_e t)}$ with  amplitude $A = 0.2D$ and frequency $f_e$. The Reynolds number $\Re = U_{\infty}D/\nu = 185$ and the Strouhal number is $S_0 = f_0 D /U_{\infty} = 0.195$, where $f_0$ is the natural vortex-shedding frequency obtained from the corresponding flow past a fixed cylinder in \cite{Schneiders2013}. The non-dimensional oscillation frequency in our simulations is $f_e/f_0 = 0.8$. The domain size is set as $[40D, 9D]$, and to obtain long-time flow diagnostics we set the downstream domain boundary in $x$ direction as an outflow boundary, as described in \citep{Chatelain2013, Gabbard2021}. The resolution is $N^* = D/h = 64$. 
The simulation is run until a dimensionless time $t^* = U_{\infty}t/D = 96.15$, before which point the force coefficients $C_d = 2F_x  /(\rho_f U_{\infty}^2D)$ and $C_l = 2F_y  /(\rho_f U_{\infty}^2D)$ reach a periodic state. The average and root-mean-square values of the force coefficients are shown in Table \ref{tab:osciCylinder}. Present results show a good agreement with those of \citet{Guilmineau2002} and \citet{Yang2009}, who both use immersed boundary methods, and \citet{Schneiders2013}, who use a cut-cell method. We note that the smallest reported domain sizes in these works cover a five times larger area than our simulations,  emphasizing the efficiency of vorticity-based approaches for external flows, even with an outflow boundary. %

\begin{table}[htb]
    \centering
    \begin{tabular}{ | c | c | c | c | c | } 
    \hline
    Author  & $\Bar{C}_d$ & $C_d'$ & $C_l'$ \\ 
    \hline\hline
    Present & 1.260 & 0.046 & 0.085\\
    \hline
    \citet{Guilmineau2002}& 1.195 & 0.036 & 0.080\\
    \hline
    \citet{Yang2009}& 1.281 & 0.042 & 0.076\\
    \hline
    \citet{Schneiders2013} & 1.279 & 0.042 & 0.082\\
    \hline
    \end{tabular}
    \caption{Long-time statistics of the lift and drag forces on an oscillating cylinder with $f_e/f_0 = 0.8$, where $\Bar{C}_d$ is the mean of drag coefficient, $C_d'$ is the root mean square of drag coefficient and $C_l'$ is the root mean square of lift coefficient.}
    \label{tab:osciCylinder}
\end{table}

\begin{figure}
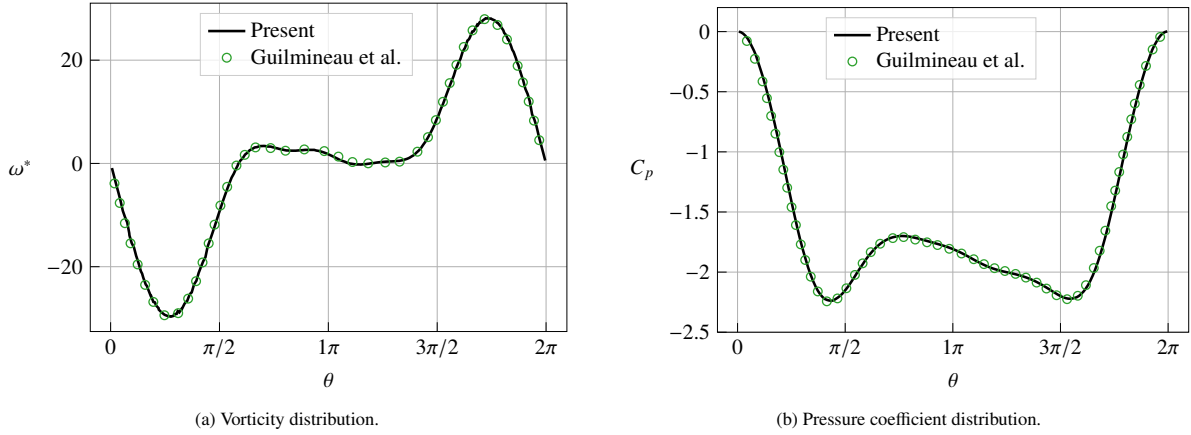

    \centering
    \begin{subfigure}{0.46\linewidth}
        \centering
        \resizebox{\textwidth}{!}{\input{tikzfigures/osciCylinder_vort}}
        \caption{Vorticity distribution.}
    \end{subfigure}
    \hspace{0.03\textwidth}
    \begin{subfigure}{0.46\linewidth}
        \centering
        \resizebox{\textwidth}{!}{\input{tikzfigures/osciCylinder_cp}}
        \caption{Pressure coefficient distribution.}
    \end{subfigure}
    \caption{Vorticity and pressure distribution on the oscillating cylinder when it reaches the top-dead end. $\theta$ is the angle measured clockwise from the stagnation point.}
    \label{fig:pressure}
\end{figure}

When the cylinder reaches its maximum $y$-direction displacement in the periodic state, in this case $t^* = 96.15$,  we compare the vorticity and pressure distributions around the surface of the cylinder with those from \citet{Guilmineau2002}. Here we define the pressure at the most upstream point of the cylinder as $p_0$, and measure the pressure distribution relative to this value. Figure~\ref{fig:pressure} shows the non-dimensional vorticity distribution $\omega^* = \omega D/U_{\infty}$, and the non-dimensional pressure distribution $C_p = 2(p - p_0)/(\rho_f U_{\infty}^2)$, which agree very well with the reference curves.

\section{Two-way coupled vorticity-velocity based Navier-Stokes simulations}\label{section:NS_twoWay}
 
\ReviewerTwo{In this section we extend our 2D Navier-Stokes solver towards two-way coupling problems of rigid bodies. We describe our approach to resolve the momentum balance with coupled bodies, then show its validation and convergence results through different test cases.} 

\subsection{Momentum balance}\label{subsection:momentum}
When the body kinematics are partially or fully governed by fluid forces, the two-way coupling between flow and body needs to be addressed. In this section we describe our approach to resolve the momentum balance with coupled bodies based on the control volume method in section~\ref{subsection:Force}.
For notational convenience, we leave out the subscript $k$ associated with the different $N_b$ bodies, as each body can be treated independently. To further simplify the notation, we rewrite the linear momentum balance \eqref{eq:fluidForce} and angular momentum balance for any given control volume as 
\begin{align}
    & \vb{F}_f = \rho_f\left(-\dv{}{t}\vb{I} + \vb{A}\right), \label{eq:simpleForce}\\
    & M_f = M_{f,0} + \vb{F}_f \times \vb{x}_c = \rho_f\left(-\dv{}{t}I_m + A_m\right).\label{eq:simpleMoment}
\end{align}
Here the new vectors $\vb{I}$ and $\vb{A}$ are defined as
\begin{align}
    &\vb{I} \equiv \int_{R\backslash B} \vb{u} \dd{A} + \oint_{\partial R} \vb{x} \times (\vu{n} \times \vb{u}) \dd{s}, \label{eq:vectorI}\\
    &\vb{A} \equiv  \oint_{\partial R} \vu{n} \cdot \bm{\gamma} \dd{s},
\end{align}
whereas the scalars $I_m$ and $A_m$ are defined as
\begin{align}
    &I_{m} \equiv \int_{R\backslash B} (\vb{x}-\vb{x}_c) \times \vb{u} \dd{A} - \oint_{\partial R} \frac{\abs{\vb{x}}^2}{2} \vu{n} \times  \vb{u} \dd{s} - \vb{x}_c \times \oint_{\partial R} \vb{x} \times (\vu{n} \times  \vb{u}) \dd{s}, \\
    &A_{m} \equiv  \oint_{\partial R} \bm{\lambda} \dd{s} - \vb{x}_c \times \oint_{\partial R} \vu{n} \cdot \bm{\gamma} \dd{s} - \vb{u}_b \times \vb{I}.\label{eq:scalarA}
\end{align}
Note that for the angular momentum equation~\eqref{eq:simpleMoment}, we explicitly shift the moment around the origin $M_{f,0}$ as defined in equation~\eqref{eq:fluidMoment_origin} to the moment around the body's center of rotation $\vb{x}_c$, whose position generally varies in time.

For a coupled body with constant volume and density, the linear and angular velocity components are governed by Newton's second law:

\begin{equation}
    \rho_b V_b \dv{\vb{u}_b}{t} = \vb{F}_f + \vb{F}_{e}, \quad \quad \rho_b I_b \dv{\Omega_b}{t} = M_f + M_e
\end{equation}
where $\bm{F}_{e}$ and $M_e$ are the external forces and torques acting on the body, $\rho_b$ is the body density, $V_b$ is the body volume, $I_b$ is the body area moment of inertia, and $\Omega_b$ is the body angular velocity.  We can combine these equations with equations \eqref{eq:simpleForce} and \eqref{eq:simpleMoment} to obtain

\begin{equation}
    \dv{}{t}\left(\frac{\rho_b V_b}{\rho_f}\vb{u}_b + \vb I\right) = \vb{A} + \frac{1}{\rho_f}\vb{F}_{e}, \quad \quad \dv{}{t}\left(\frac{\rho_b I_b}{\rho_f}\Omega_b + I_m\right) = A_m + \frac{1}{\rho_f}M_{e}
\end{equation}
where all time derivatives are collected on the left-hand side.  For convenience, we define 
\begin{equation}
    \vb{L} \equiv \frac{\rho_b V_b}{\rho_f}\vb{u}_b + \vb I, \quad \quad L_m \equiv \frac{\rho_b I_b}{\rho_f}\Omega_b + I_m,
\end{equation}
thus we get
\begin{equation}
\label{eq:basic2way}
    \dv{\vb{L}}{t} = \vb{A} + \frac{1}{\rho_f}\vb{F}_{e}, \quad \quad 
    \dv{L_m}{t} = A_m + \frac{1}{\rho_f}M_{e}.
\end{equation}
These equations represent linear and angular momentum balance on a control volume with an immersed, coupled body, and must be integrated in time alongside the vorticity-velocity Navier-Stokes equations~\eqref{eq:transport} and \eqref{eq:kelvin}. 

To do so, we opt for a simple weak coupling approach using an explicit, second- or third-order accurate Runge-Kutta integration for all time derivatives. In any given Runge-Kutta substep, starting from a vorticity field; a set of $N_b$ linear and angular body positions and velocities; and a set of $N_b$ circulation constraints, we perform the following steps. \\

\noindent \textit{Step 1: Velocity reconstruction.} 
The velocity reconstruction is performed as described \ReviewerTwo{in section~\ref{subsection:NS_fixed}} \ReviewerTwoStrike{above and in \cite{Gabbard2021}}. This leads to a unique velocity field satisfying the appropriate far-field boundary conditions, as well as the no-through boundary condition on each body. \\

\noindent \textit{Step 2: Right-hand side computation.} After the velocity reconstruction we compute for each of the $N_b$ control volumes the vectors $\vb{I}$ and $\vb{A}$, and the scalars $I_m$ and $A_m$, according to the equations from (\ref{eq:vectorI}) to (\ref{eq:scalarA}). Further, we compute at all grid points the right-hand side  for the vorticity transport equation, as well as the right-hand sides for each of the $N_b$ circulations $\Gamma_k$ \ReviewerTwo{as shown in section~\ref{subsection:NS_fixed}} \ReviewerTwoStrike{\cite{Gabbard2021}}. \\

\noindent \textit{Step 3: Time integration.} Next, we integrate the vorticity transport equation, the body positions, and the quantities $\vb{L}$ and $L_m$, and the circulation constraints in time using the appropriate Runge-Kutta substep algorithm. \\

\noindent \textit{Step 4: Body velocity update.} Before advancing to the next timestep, we need to extract the body linear and angular velocities from the updated quantities $\vb{L}$ and $L_m$, respectively. Denoting values that have been advanced in time with superscript $n+1$, we are faced with the following equation for the linear velocities 
\begin{equation}
    \vb{L}^{n+1} = \frac{\rho_b V_b}{\rho_f}\vb{u}_b^{n+1} + \vb{I}^{n+1},
    \label{eq:basic-2way}
\end{equation}
where the left-hand side is known from the Runge-Kutta time integration, but $\vb{u}_b^{n+1}$ and $\vb{I}^{n+1}$ are both unknown because they haven't been explicitly updated. Further, since $\vb{I}^{n+1}$ depends on the velocity field $\vb{u}^{n+1}$, which in turn depends on $\vb{u}_b^{n+1}$ through the boundary conditions in the velocity reconstruction step, the right-hand side is in fact an implicit expression in $\vb{u}_b^{n+1}$. Here we make the simple approximation $\vb{I}^{n+1} \approx \vb{I}^n$ to get an explicit expression for the new body velocities. For a time step size of $\Delta t$, this approximation incurs a first-order time integration error:
\begin{equation}
    \vb{u}_b^{n+1} = \frac{\rho_f }{\rho_b V_b} \left(\vb{L}^{n+1} - \vb{I}^n  \right) + \frac{1}{V_b} \frac{\rho_f}{\rho_b} \left(\mathcal{O}(\Delta t) \right).
    \label{eq:two-way-coupling}
\end{equation}
An analogous assumption is made for the angular momentum, leading to a similar first-order accurate temporal discretization error in the angular velocity.

\ReviewerThree{We note that our approach in equation~\eqref{eq:two-way-coupling} can be interpreted as a first step within an iterative strong coupling algorithm, as $\vb{I}^{n+1}$ relies on $\vb{u}_b^{n+1}$ through the velocity field $\vb{u}^{n+1}$. Following a simple fixed-point iteration scheme, one can use an updated $\vb{u}_b$ to recompute the velocity field $\vb{u}$ using Step~1, then execute Steps 2--4 to find an updated body velocity $\vb{u}_b$, and repeat this until equation~\eqref{eq:basic-2way} is satisfied. We show below  (section~\ref{subsection:LambOseen}) that this indeed leads to second-order temporal convergence even at low density ratios. However, in this naive formulation each iteration involves solving a global elliptic equation, leading to a prohibitive computational cost in practice. We therefore rely on equation~\eqref{eq:two-way-coupling} in the rest of this work, instead focusing on analyzing the consequence of the first-order temporal error in practical simulations below.}

\subsection{Two-way coupling results}\label{section:results_twoWay}

In this section we demonstrate the ability of the two-way coupling algorithm on a set of test problems where the obstacle motion is driven both by fluid forces and external forces.

\subsubsection{Convergence: Forced Lamb-Oseen vortex}
\label{subsection:LambOseen}
To investigate the convergence properties of the two-way coupling algorithm, we construct a test case based on the analytic Lamb-Oseen vortex flow. The Lamb-Oseen vortex is an axisymmetric Gaussian patch of vorticity with circulation $\Gamma$ that diffuses over time in an unbounded domain:
\begin{equation}
\omega_{LO}(\vb{x}, t) = \frac{\Gamma}{4 \pi \nu t} e^{-r^2/(4\nu t)},
\label{eq:LOomegafield}
\end{equation}
where $r = \| \vb{x} - \vb{x}_c\|$. In previous work \cite{Gabbard2021}, we initialize a simulation with a Lamb-Oseen vortex evaluated at some time $t>0$, and place a solid cylinder of radius $R$ at the center of the vortex. By setting the time-dependent angular velocity of the cylinder to 
\begin{equation}
\Omega_b(t) = \frac{\Gamma}{2\pi R^2}\left(1 - e^{-\frac{R^2}{4t\nu}}\right),
\label{eq:LOomega}
\end{equation}
the rotational velocity at the cylinder boundary matches that of the exact solution. Consequently, the Lamb-Oseen vortex field remains an exact solution to the flow field outside of the cylinder and can thus be used for error analysis. Here this approach is revisited as a two-way coupling problem. Since the intersections of this geometry with the Cartesian grid do not change over time, this case is especially useful to analyze errors associated with the momentum balance in isolation from  errors associated with body motion. 

To establish this test case as a two-way coupling problem, we use the exact solution to the Lamb-Oseen vortex to find the exact angular acceleration $\alpha_b(t)$ as well as the hydrodynamic torque $M_f(t)$ on a cylinder of radius $R$ placed at the center of the vortex:
\begin{align}
    \begin{split}
    \alpha_b(t) &= -\frac{\Gamma}{8\pi t^2 \nu}e^{-\frac{R^2}{4t\nu}}, \\
    M_f(t) &= -\ReviewerTwo{\Gamma \nu} \left(2 \ReviewerTwo{\rho_f} \ReviewerTwo{-} \frac{\ReviewerTwo{\rho_f}(R^2+4t\nu)}{2t \ReviewerTwo{\nu}}e^{-\frac{R^2}{4t\nu}}\right).
    \end{split}
\end{align}
From Newton's second law we know that $\rho_b I_b\alpha_b = M_f + M_e$, where $I_b = \pi R^4/2$ is the area moment of inertia of the cylinder, so that the external torque $M_e$ must be
\begin{equation}
M_e(t) = \Gamma \ReviewerTwo{\nu} \left[ 2 \rho_f -  \left( \rho_b \frac{R^4}{16 t^2 \nu\ReviewerTwo{^2}} + \rho_f \frac{R^2+4t\nu}{2 t \ReviewerTwo{\nu}} \right)e^{-\frac{R^2}{4t\nu}} \right].
\end{equation}
Consequently, imposing $M_e(t)$ as an external torque in the two-way coupling approach should lead to the correct angular acceleration of the cylinder.

We simulate this case with the settings in Table \ref{tab:LambOseen}, using a Reynolds number of $Re = 1000 \pi \approx 3142$. The resulting angular velocity $\Omega_b$ is normalized to $\Omega_b^* = \Omega_b(2\pi R^2)/\Gamma$, and we use the $L_2$ and $L_\infty$ norms of the error in the time evolution of $\Omega_b^*(t)$ between times $t_0$ and $t_f$ as error metrics. We measure the convergence of these error metrics independently with spatial and temporal resolution refinements. For spatial refinement, simulations are run at a constant CFL-based timestep criterion, so that $\Delta t \sim  h$, and errors are computed with respect to the exact solution for resolutions $N^* = 11$ to $N^* = 683$, with $N^* = D/h$. For temporal refinement, we fix the resolution to $N^* = 11$, the density ratio to $\rho_b/\rho_f = 0.4$, and reduce the timestep $\Delta t^* = \Delta t \Gamma/(2\pi R^2)$ from $7.1 \times  10^{-2}$ to $2.2 \times 10^{-3}$. Here we use the results from an even smaller timestep, $\Delta t^* = 2.2 \times  10^{-4}$, as the reference value for the error computation. 

\begin{table}[htb]
    \centering
    \begin{tabular}{ | c | c |} 
    \hline
    Element & Parameters  \\ 
    \hline\hline
    Grid & $\vb{x}_{ij} = (0.9 i/N_x, 0.9 j/N_y)$ for $0 \leq i \leq N_x - 1$, $0 \leq j \leq N_y - 1$. $N_x = N_y$\\
    \hline
    Time & $t_0 = 3.0$, $t_f = 3.5$\\
    \hline
    Obstacle & $\vb{x}_c = (0.457, 0.457)$, $R = 0.15$, $D = 2R$\\
    \hline
    Flow & $\Gamma = \pi$, $\nu = 0.001$\\
    \hline
    \end{tabular}
    \caption{Simulation parameters for the two-way coupling Lamb-Oseen case, where $N_x = 0.9N^*/D$ with $N^* = D/h$ is the number of grid points along the diameter.}
    \label{tab:LambOseen}
\end{table}

\begin{figure}
    \centering
    \begin{subfigure}{0.46\linewidth}
        \centering
        \resizebox{\textwidth}{!}{\input{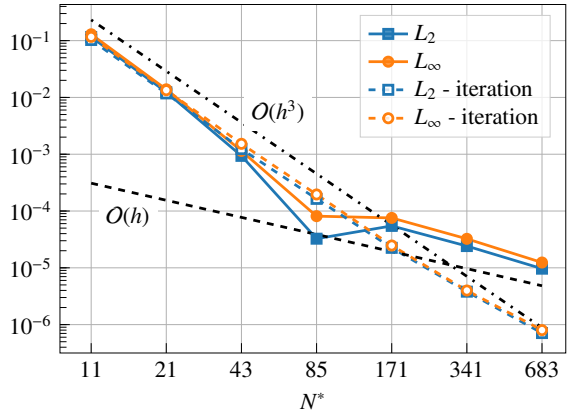}}
        \caption{Spatial convergence of $\rho_b/\rho_f = 0.1$.}
        \label{subfigure:rho0.1}
    \end{subfigure}        
    \hspace{0.03\textwidth}    
    \begin{subfigure}{0.46\linewidth}
        \centering
        \resizebox{\textwidth}{!}{\input{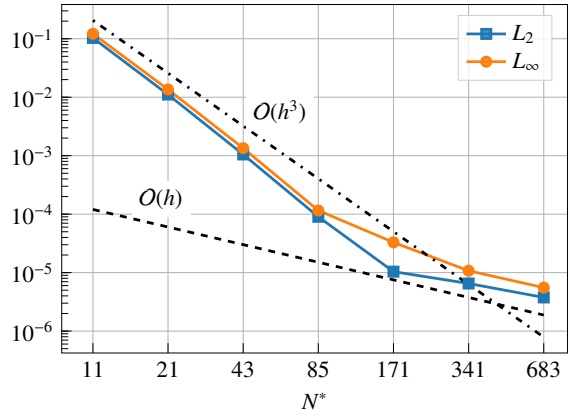}}
        \caption{Spatial convergence of $\rho_b/\rho_f = 0.2$.}
        \label{subfigure:rho0.2}
    \end{subfigure}
    \par\vspace{20pt}
    \begin{subfigure}{0.46\linewidth}
        \centering
        \resizebox{\textwidth}{!}{\input{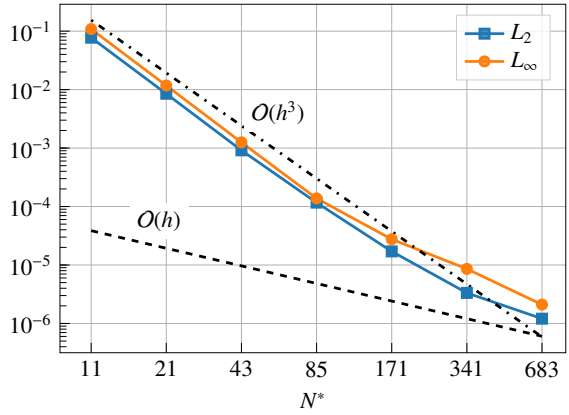}}
        \caption{Spatial convergence of $\rho_b/\rho_f= 0.4$.}
        \label{subfigure:rho0.4}
    \end{subfigure}
    \hspace{0.03\textwidth}    
    \begin{subfigure}{0.46\linewidth}
        \centering
        \resizebox{\textwidth}{!}{\input{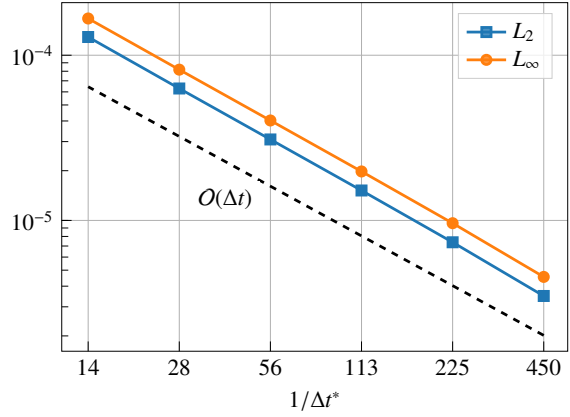}}
        \caption{Temporal convergence of resolution $N^* = 11$.}
        \label{subfigure:temporal}
    \end{subfigure}
    \caption{Spatial and temporal convergence of a two-way coupling cylinder in Lamb-Oseen vortex. }
    \label{fig:cylresults}
\end{figure}

Figure~\ref{fig:cylresults}(a)--(c) shows the spatial convergence at density ratios of $\rho_b/\rho_f \in [0.1, 0.2, 0.4]$. For small values of $N^*$, we observe a third-order convergence consistent with the behavior of our algorithm for convection-dominated flows \cite{Gabbard2021}. When $N^*$ increases, the convergence plot shows a second regime where the spatial error behaves as $\mathcal{O}(h)$. Since here $\Delta t \sim \Delta h$, this error is associated with the first-order temporal error introduced in our two-way coupling algorithm, as discussed in section~{\ref{subsection:momentum}}. 
As argued there, the temporal error is divided by the density ratio $\rho_b/\rho_f$, so that we expect the error values to decay as the density ratio increases. This behavior is demonstrated in panels~(a) -- (c) in Figure~\ref{fig:cylresults}: as $\rho_b/\rho_f$ increases, the prefactor for the first-order error term decreases. Consequently, the transition from third- to first-order convergence occurs at higher values of $N^*$ as $\rho_b/\rho_f$ increases. 

\ReviewerThree{Panel (a) also shows the spatial convergence of $\rho_b/\rho_f = 0.1$ under the iteration method discussed above in section~\ref{subsection:momentum}. For these results, the iterations are stopped when the relative error of $\Omega_b$ reaches $10^{-8}$. The convergence plot shows a third-order convergence throughout all the resolutions at the lowest density ratio tested, providing further support for the origin of the first-order temporal error term and a direction for future developments. However, as discussed in section~\ref{subsection:momentum}, the iteration method requires prohibitive computational cost, especially at high resolutions, so we focus on equation~\eqref{eq:two-way-coupling} in the rest of this work.}

Panel (d) of Figure~\ref{fig:cylresults} shows the error evolution associated with temporal refinement at fixed spatial resolution, and indicates first-order temporal convergence. Since the boundary of the cylinder remains stationary with respect to the background grid, we do not need to consider the mixed error term associated with boundary motion discussed in section~\ref{subsection:IIM_moving_1D_convergence}. The first-order temporal convergence can thus be solely attributed to the $\mathcal{O}(\Delta t)$ error in the coupling approach. 

Overall, Our error analysis for a two-way coupled problem confirms the theoretical prediction: there exists a first-order temporal error that becomes more dominant at lower density ratios. For larger density ratios the spatial error of the uncoupled algorithm dominates, in which case  the two-way coupling approach does not detract from the overall second-order convergence behavior of the free-space discretization.

\subsubsection{Sedimentation of a cylinder}
To test the two-way coupling behavior on a case with moving boundaries, we consider the sedimentation of a cylinder in a quiescent flow, both in comparison to reference results and in a convergence analysis.

\paragraph{Comparison with references}
We follow the work in \cite{Namkoong2008}, which has since been used in several other validation studies. For this case, a cylinder with radius $D$ and density ratio $\rho_b / \rho_f = 1.01$ is released from rest in a viscous flow with viscosity $\nu$. Besides the density ratio, the resulting dynamics are governed by the Archimedes number, which here is $Ar = V_g D / \nu = 138.42$ with $V_g = \sqrt{| 1 - \rho_b/\rho_f| g D}$ the gravitational velocity scale~\cite{Ern2012}. Time is normalized as $t^* = t V_g/D$, and the body velocity is normalized as $\vb{u}_b^* = \vb{u}_b/V_g$. From \cite{Namkoong2008}, the Reynolds number based on their computed terminal velocity $U$ is $Re = U D / \nu = 156.31$, associated with a relative terminal velocity $U/V_g = Re/Ar = 1.129$. 

Our simulations are run in a domain of size $[16D \times 120D]$ until $t^* = 90$ with free-space boundary conditions on all sides. We use resolutions of $N^* = D/h \in [48, 64, 96, 128]$. For $N^* = 128$ an initial perturbation is added to the cylinder to break symmetry, since the numerical error alone does not do so until late in the simulation. The perturbation consists of an imposed angular velocity of the cylinder when $t^* < 4$ of form $\Omega_b = 0.05 U\sin(0.5\pi t^*)/D$, after which the angular velocity is released to be determined by the two-way coupling algorithm.

\begin{figure}[htb]
    \centering
    \begin{subfigure}[t][][t]{0.5\textwidth}
    \resizebox{\textwidth}{!}{\input{tikzfigures/sedCylinder}}
    \caption{Normalized velocities of the falling cylinder simulated at $N^* = 96$ compared with the results from \citet{Gazzola:2011a} (Gazzola) and \citet{Namkoong2008} (Namkoong).}
    \end{subfigure}
    \hspace{0.05\textwidth}
    \begin{subfigure}[t][][t]{0.25\textwidth}
        \centering
        \includegraphics[width=1.0\textwidth]{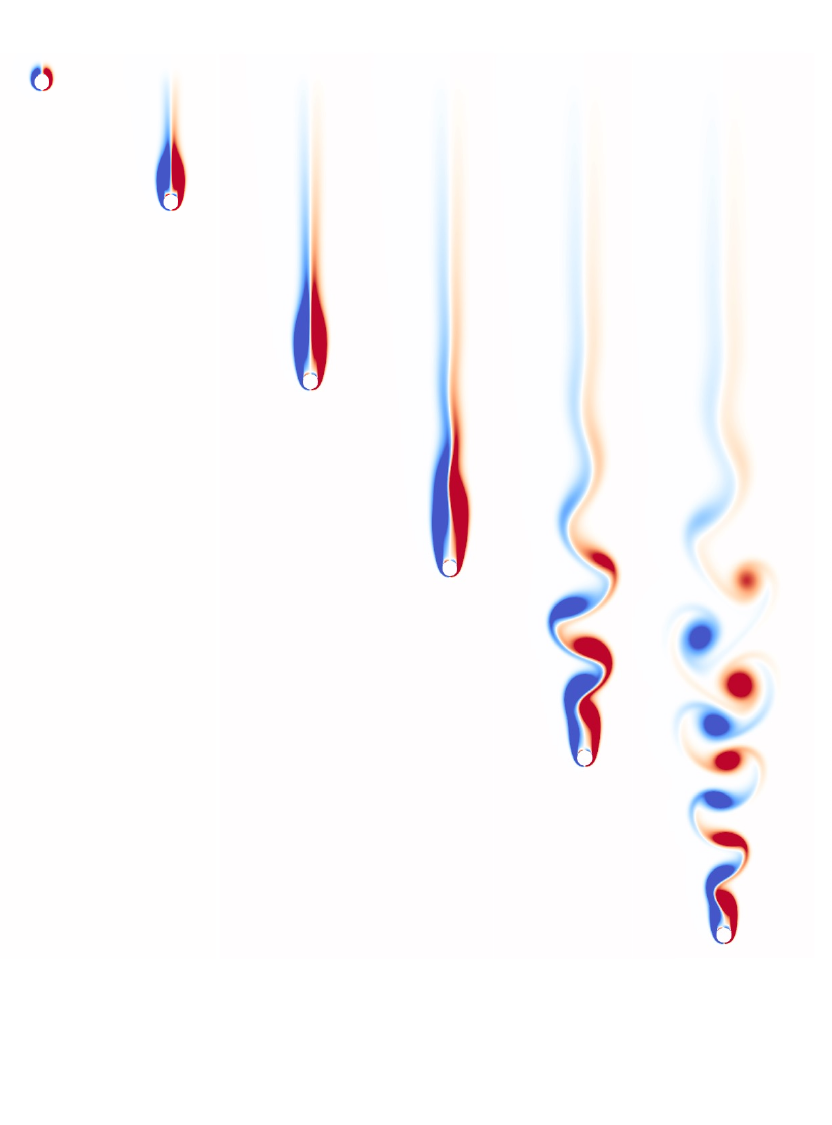}
        \caption{Vorticity field of present results at $N^* = 96$ at different normalized times: from left to right $t^* = 2.5$, $t^* = 11$, $t^* = 22$, $t^* = 33$, $t^* = 44$, $t^* = 55$.}
    \end{subfigure}
    \caption{Results for sedimentation of a cylinder at Archimedes number $Ar = 138.42$.}
    \label{fig:sedCylinder}
\end{figure}

Steady-state values of the terminal velocity as a function of spatial resolution are shown in Table~\ref{tab:sed_long}, showing that the terminal velocity at the two higher resolutions matches well (within ~1\%) with the reference.  The time variations of the linear velocities for the case  $N^* = 96$ is shown in Fig~\ref{fig:sedCylinder}, compared with the combined finite-element method results from  \cite{Namkoong2008} and the penalization-based approach in \citep{Gazzola:2011a}. At early time, we observe that our approach matches well but delays the shedding until significantly later times than the other references. Once the cylinder starts its vortex shedding, our results again match well with steady-state values in all three velocity components.

\begin{table}[htb]
    \centering
    \begin{tabular}{ | c | c | c | c | c |} 
    \hline
    $N^*$ & 48 & 64 & 96 & 128 \\ 
    \hline
    $\Bar{u}_{b,y}^*$ & 1.091 & 1.104 & 1.117 & 1.121\\
    \hline
    \end{tabular}
    \caption{Steady-state values of the y-direction body velocity with different resolutions $N^*$, where $\Bar{u}_{b,y}^*$ is the mean value of $u_{b,y}^*$. The y-direction terminal velocity from \cite{Namkoong2008} is 1.129.}
    \label{tab:sed_long}
\end{table}

\paragraph{Convergence}
To examine the convergence of our two-way coupling algorithm with a moving interface we repeat the above setup with a density ratio $\rho_b / \rho_f = 1.2$. Further, we fix the horizontal and angular degrees of freedom and conduct a convergence analysis based on the error in the time evolution of $\Delta y_c^* = \Delta y_c/D$ between $t^*_0 = 0.0$ and $t^*_f = 0.2$. The domain size in all convergence tests is $10D/3 \times 10D/3$.

\begin{figure}
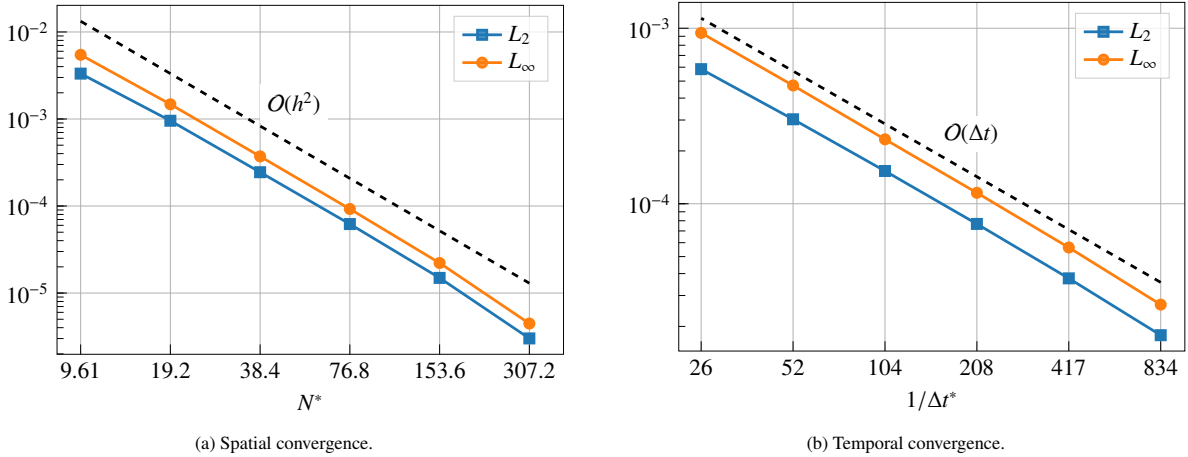

    \centering
    \begin{subfigure}{0.46\linewidth}
        \centering
        \resizebox{\textwidth}{!}{\input{tikzfigures/sedCylinder_spatial}}
        \caption{Spatial convergence.}
        \label{subfig:sedSpatial}
    \end{subfigure}
        \hspace{0.03\textwidth}
        \begin{subfigure}{0.46\linewidth}
        \centering
        \resizebox{\textwidth}{!}{\input{tikzfigures/sedCylinder_temporal}}
        \caption{Temporal convergence.}
        \label{subfig:sedTemporal}
    \end{subfigure}        
    \caption{Convergence plot of $\Delta y_{c}^*$ of a falling cylinder. }
    \label{fig:sedConvergence}
\end{figure}

For spatial convergence, we vary $N^* = D/h$ in the range $N^* = 9.61$ to $N^* = 307.2$, and use a result from $N^* = 614.4$ as reference. We use a CFL-based timestep constraint so that $\Delta t \sim h$ throughout. Figure \ref{subfig:sedSpatial} shows a second-order spatial convergence rate, demonstrating that at this density ratio and resolution range, the first-order coupling error does not dominate. This underscores that in many practical scenarios of interest, our algorithm for two-way coupling with moving boundaries is effectively second-order accurate.

For temporal convergence, the spatial resolution is fixed at $N^* = 9.6$, and different time step sizes $\Delta t^*$ from $3.8 \times  10^{-2}$ to $1.2\times 10^{-3}$ are tested. The result using $\Delta t^* = 1.2 \times  10^{-4}$ is used as a reference for the error computation. Figure~\ref{subfig:sedTemporal} shows that the temporal error has a first-order convergence rate arising from our two-way coupling approach.

\subsubsection{Single elastically mounted cylinder}
Our last two test cases concern the long-time diagnostics of  elastically mounted cylinders in a free-stream flow. First, we consider a single cylinder of mass $m$ and diameter $D$ mounted on a spring with linear stiffness constant $k$. The cylinder is allowed to move in the $y$ direction, forced by an incoming flow with velocity $U_\infty$ in the $x$-direction. The $x$-displacement and angular displacement of the cylinder are fixed at all times. We follow the setup considered in \cite{Ahn2006} and further used in \cite{Bao2012} with Reynolds number $Re = U_{\infty}D/\nu = 150$, reduced velocity $U_{red}^2 = 8\pi^2 U_{\infty}^2 \rho_f/k$, and reduced mass $M_{red} = \pi/2 (\rho_b/\rho_f) = 4$. Time is non-dimensionalized as $t^* = U_{\infty}t/D$.

\begin{figure}
    \centering
    \includegraphics[width=\textwidth]{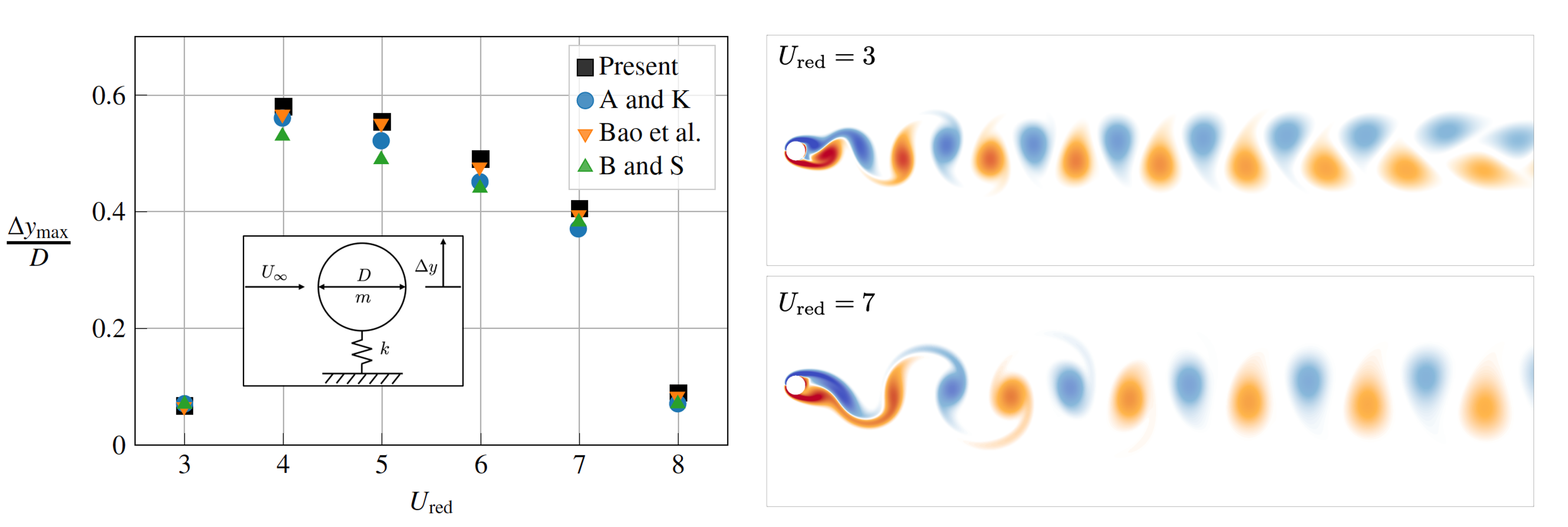}
    \caption{Left: Maximum amplitude response versus the reduced velocity compared with \citet{Ahn2006} (A and K), \citet{Bao2012} and \citet{BandS2009} (B and S), where $Re = 150$ and $M_{red} = 4$. The inset shows the numerical setup. Right: Vorticity field visualized at $t^* = 105$ for $U_{red} = 3$ (top right) and $U_{red} = 7$ (bottom right). }
    \label{fig:single_mountedCylinder}
\end{figure}

Our simulation domain is set as $[12D, 40D]$ with resolution $N^* = D/h = 64$. The downstream domain boundary set as an outflow boundary.
Initially, for $U_{red}$ from 4 to 7, the cylinder is set with an initial displacement $\Delta y/D = 0.1$ to trigger the vortex shedding. We simulate cases with $U_{red}$ increasing from $3$ to $8$ and report the maximum amplitude of each case in figure \ref{fig:single_mountedCylinder}. All the simulations are run until $t^* = 105$. Our results match very well with the references in \cite{Ahn2006,Bao2012}, who both used an Arbitrary-Lagrangian-Eulerian (ALE) formulation.

\subsubsection{Tandem elastically mounted cylinders}
\begin{figure}
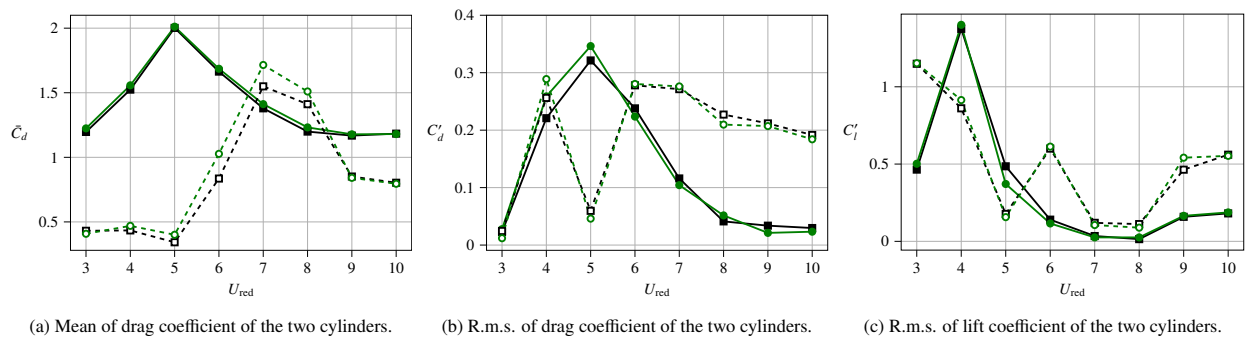

    \begin{subfigure}[b]{0.33\linewidth}
        \centering
        \resizebox{\textwidth}{!}{\input{tikzfigures/cd_mean}}
        \caption{Mean of drag coefficient of the two cylinders.}
    \end{subfigure}
    \begin{subfigure}[b]{0.33\linewidth}
        \centering
        \resizebox{\textwidth}{!}{\input{tikzfigures/cd_rms}}
        \caption{R.m.s. of drag coefficient of the two cylinders.}
    \end{subfigure}
    \begin{subfigure}[b]{0.33\linewidth}
        \centering
        \resizebox{\textwidth}{!}{\input{tikzfigures/cl_rms}}
        \caption{R.m.s. of lift coefficient of the two cylinders.}
    \end{subfigure}
    \caption{Configuration of the tandem-mounted cylinder and drag and lift force coefficient versus different reduced velocities, where $Re = 150$ and $M_{red} = 4$. Present results (black lines/squares) are compared with results from \citet{Bao2012} (green lines/circles). In all plots, solid lines correspond to the upstream cylinder and dashed lines correspond to the downstream cylinder.}
    \label{fig: 2mountedCylinders}
\end{figure}

To investigate the performance of our approach with multiple moving obstacles in a two-way coupling setting, we consider  tandem spring-mounted cylinders in a free-stream flow. Here each cylinder is connected to springs with stiffness $k$ in both $x$ and $y$ directions. At rest, the cylinders are spaced $5D$ apart in the $x$-direction, and aligned in the $y$-direction.
The cylinders are released to move in both $x$ and $y$ directions, with the angular velocity kept to zero at all times. The domain is chosen as $[16D, 60D]$, and as for the single spring-mounted cylinder, the downstream $x$ boundary is treated as an outflow plane. Vortex shedding is triggered by initializing both cylinders with a vertical displacement of $\Delta y/D = 0.05$, and all the simulations are run until $t^* = 300$. All other settings for the simulations of tandem-mounted cylinders are the same as the single-mounted cylinder considered above.  Figure~\ref{fig: 2mountedCylinders} shows for each cylinder the mean and root-mean-square value of the drag coefficient $C_d = 2F_x  /(\rho_f U_{\infty}^2 D)$ as well as the root-mean-square value of the lift coefficient $C_l = 2F_y  /(\rho_f U_{\infty}^2 D)$ for eight different values of $U_{red}$. We compare the results with those reported in \cite{Bao2012} computed using an Arbitrarily-Lagrangian-Eulerian method, and observe good agreement. This demonstrates that our algorithm is capable of handling multiple moving bodies in a two-way coupled setting with high-fidelity.

\section{Conclusion}\label{section:conclusion}
In this work we presented a 2D vorticity-velocity Navier-Stokes solver with sharp boundary treatment that can handle one- and two-way coupled bodies. To achieve this, we have derived and analyzed a treatment for moving sharp interfaces that can be combined easily with explicit Runge-Kutta type time integrators and does not require knowledge of a boundary condition on the solution. Our numerical experiments in 1D and 2D have shown that the scheme leads to a mixed spatio-temporal error term that does not deteriorate the convergence of the algorithm under a fixed CFL condition. 

Based on this moving sharp interface treatment, we extended a stationary-body IIM vorticity-velocity solver to handle moving bodies, and provided a range of validation and convergence tests for one-way coupled problems. Notably, we simulated an impulsively started cylinder using a body-based and flow-based frame of reference, and observed that the discretization errors do not significantly change between simulations where the boundary is moving relative to the grid compared to those where the boundary is stationary. Further, we compared our approach with a first-order penalization scheme and demonstrated that, to achieve a maximum error of about 1\% compared to a high-resolution reference solution, our IIM approach requires about 20 times less grid points than the penalization method. 

Finally, we extended our one-way coupled Navier-Stokes solver to two-way coupled problems by means of a control volume method. The approach does not require any pressure computation, and our validations demonstrate that is suitably reproduces reference results on cylinder sedimentation and spring-mounted cylinders. Our weak coupling strategy leads to a first-order temporal error term that becomes dominant when considering small density-ratio bodies; for large density ratios, our two-way coupling approach effectively converges with second order convergence under constant CFL. 

Beyond these contributions, our methodology carries over some advantageous features of our previous work \cite{Gabbard2021}. First, the sharp-interface treatment enables the computation of the vorticity flux on immersed boundaries, which, when integrated, yields a pressure distribution over the body surface in a computationally inexpensive post-processing step. Second, the use of conservative finite differences leads to exact discrete conservation of circulation in any grid-aligned control region, which provides a robust way to enforce Kelvin's circulation theory around multiple obstacles embedded in the same domain. Third, the outflow condition devised in \cite{Chatelain2013} and further tested in \cite{Gabbard2021} enables long-time simulations in compact domains. Together, this leads to an efficient and accurate solver for rigid-body flow-structure interaction simulations.

Future work will address three possible extensions of this work. First, we consider improving the two-way coupling strategy to a strong coupling approach. This will eliminate the first-order error term and make the algorithm second-order for all density ratios. It will further improve stability of our approach for small density ratios, as the added mass term will be treated implicitly. Since the term in the two-way coupling that is current treated weakly is linear in the flow variables, a strong coupling strategy can be envisioned using fast linear solvers. Second, we consider the extension from rigid bodies to those with deforming boundaries. For one-way coupled problems with deforming boundaries (e.g.\ self-propelled swimmers with imposed kinematics), the current streamfunction-based velocity reconstruction needs to be amended with a velocity potential. For two-way coupled problems with deforming bodies (e.g.\ flows interacting with bulk elastic structures), the algorithm further needs to use surface integral force distributions instead of our current control volume approach. Though our simulations indicate this approach is viable, high-fidelity force distributions would likely require higher resolutions than high-fidelity force integrals \cite{Gabbard2021}. Third, we consider the extension to 3D problems. Here the use of high-order immersed geometry discretizations is likely to significantly reduce the computational cost of high-fidelity simulations, especially in combination with high-order multiresolution grid adaptation techniques \cite{Gillis:2022}. On the other hand, the use of the vorticity-velocity formulation comes with significant challenges, such as maintaining the solenoidal nature of the vorticity field, as well as costs, such as solving a vector Poisson equation each timestep. Extending the IIM approach described here to 3D, high order simulations for multiphysics flow problems is currently an active area of research in our group.

\section*{Acknowledgements}
We wish to acknowledge financial support from two MathWorks Engineering Fellowships (XJ and JG); from MIT Sea Grant (WVR, XJ); and from an Early Career Award from the Department of Energy, Program Manager Dr.~Steven~Lee, award number DE-SC0020998 (JG, WVR).

\bibliographystyle{model1-num-names}
\bibliography{refs}

\begin{thebibliography}{57}
\expandafter\ifx\csname natexlab\endcsname\relax\def\natexlab#1{#1}\fi
\providecommand{\url}[1]{\texttt{#1}}
\providecommand{\href}[2]{#2}
\providecommand{\path}[1]{#1}
\providecommand{\DOIprefix}{doi:}
\providecommand{\ArXivprefix}{arXiv:}
\providecommand{\URLprefix}{URL: }
\providecommand{\Pubmedprefix}{pmid:}
\providecommand{\doi}[1]{\href{http://dx.doi.org/#1}{\path{#1}}}
\providecommand{\Pubmed}[1]{\href{pmid:#1}{\path{#1}}}
\providecommand{\bibinfo}[2]{#2}
\ifx\xfnm\relax \def\xfnm[#1]{\unskip,\space#1}\fi
\bibitem[{Griffith and Patankar(2020)}]{griffith2020immersed}
\bibinfo{author}{B.~E. Griffith}, \bibinfo{author}{N.~A. Patankar},
\newblock \bibinfo{title}{Immersed methods for fluid--structure interaction},
\newblock \bibinfo{journal}{Annual review of fluid mechanics}
  \bibinfo{volume}{52} (\bibinfo{year}{2020}) \bibinfo{pages}{421--448}.
\bibitem[{Mittal and Iaccarino(2005)}]{mittal2005}
\bibinfo{author}{R.~Mittal}, \bibinfo{author}{G.~Iaccarino},
\newblock \bibinfo{title}{Immersed boundary methods},
\newblock \bibinfo{journal}{Annu. Rev. Fluid Mech.} \bibinfo{volume}{37}
  (\bibinfo{year}{2005}) \bibinfo{pages}{239--261}.
\bibitem[{Eldredge(2007)}]{Eldredge2007}
\bibinfo{author}{J.~D. Eldredge},
\newblock \bibinfo{title}{{Numerical simulation of the fluid dynamics of 2D
  rigid body motion with the vortex particle method}},
\newblock \bibinfo{journal}{Journal of Computational Physics}
  \bibinfo{volume}{221} (\bibinfo{year}{2007}) \bibinfo{pages}{626--648}.
\bibitem[{Eldredge(2008)}]{Eldredge2008}
\bibinfo{author}{J.~D. Eldredge},
\newblock \bibinfo{title}{{Dynamically coupled fluid–body interactions in
  vorticity-based numerical simulations}},
\newblock \bibinfo{journal}{Journal of Computational Physics}
  \bibinfo{volume}{227} (\bibinfo{year}{2008}) \bibinfo{pages}{9170--9194}.
\bibitem[{Calhoun(2002)}]{Calhoun2002}
\bibinfo{author}{D.~Calhoun},
\newblock \bibinfo{title}{{A Cartesian Grid Method for Solving the
  Two-Dimensional Streamfunction-Vorticity Equations in Irregular Regions}},
\newblock \bibinfo{journal}{Journal of Computational Physics}
  \bibinfo{volume}{176} (\bibinfo{year}{2002}) \bibinfo{pages}{231 -- 275}.
\bibitem[{Li and Wang(2003)}]{Li2003}
\bibinfo{author}{Z.~Li}, \bibinfo{author}{C.~Wang},
\newblock \bibinfo{title}{{A fast finite difference method for solving
  Navier-Stokes equations on irregular domains}},
\newblock \bibinfo{journal}{Communications in Mathematical Sciences}
  \bibinfo{volume}{1} (\bibinfo{year}{2003}) \bibinfo{pages}{180 -- 196}.
\bibitem[{Russell and Wang(2003)}]{Russell2003}
\bibinfo{author}{D.~Russell}, \bibinfo{author}{Z.~J. Wang},
\newblock \bibinfo{title}{{A cartesian grid method for modeling multiple moving
  objects in 2D incompressible viscous flow}},
\newblock \bibinfo{journal}{Journal of Computational Physics}
  \bibinfo{volume}{191} (\bibinfo{year}{2003}) \bibinfo{pages}{177--205}.
\bibitem[{Linnick and Fasel(2005)}]{Linnick2005}
\bibinfo{author}{M.~N. Linnick}, \bibinfo{author}{H.~F. Fasel},
\newblock \bibinfo{title}{{A high-order immersed interface method for
  simulating unsteady incompressible flows on irregular domains}},
\newblock \bibinfo{journal}{Journal of Computational Physics}
  \bibinfo{volume}{204} (\bibinfo{year}{2005}) \bibinfo{pages}{157 -- 192}.
\bibitem[{Singhal and Kalita(2022)}]{Singhal2022}
\bibinfo{author}{R.~Singhal}, \bibinfo{author}{J.~C. Kalita},
\newblock \bibinfo{title}{{An efficient explicit jump high-order compact
  immersed interface approach for transient incompressible viscous flows}},
\newblock \bibinfo{journal}{Physics of Fluids} \bibinfo{volume}{34}
  (\bibinfo{year}{2022}).
\bibitem[{Marichal et~al.(2016)Marichal, P.Chatelain, and
  G.Winckelmans}]{Marichal2016}
\bibinfo{author}{Y.~Marichal}, \bibinfo{author}{P.Chatelain},
  \bibinfo{author}{G.Winckelmans},
\newblock \bibinfo{title}{{Immersed interface interpolation schemes for
  particle–mesh methods}},
\newblock \bibinfo{journal}{Journal of Computational Physics}
  \bibinfo{volume}{326} (\bibinfo{year}{2016}) \bibinfo{pages}{947 -- 972}.
\bibitem[{Marichal(2014)}]{MarichalThesis}
\bibinfo{author}{Y.~Marichal}, \bibinfo{title}{{An immersed interface vortex
  particle-mesh method}}, Ph.D. thesis, UCLouvain, \bibinfo{year}{2014}.
\bibitem[{Gillis et~al.(2019)Gillis, Marichal, Winckelmans, and
  Chatelain}]{Gillis2019}
\bibinfo{author}{T.~Gillis}, \bibinfo{author}{Y.~Marichal},
  \bibinfo{author}{G.~Winckelmans}, \bibinfo{author}{P.~Chatelain},
\newblock \bibinfo{title}{A 2d immersed interface vortex particle-mesh method},
\newblock \bibinfo{journal}{Journal of Computational Physics}
  \bibinfo{volume}{394} (\bibinfo{year}{2019}) \bibinfo{pages}{700 -- 718}.
\bibitem[{Gabbard et~al.(2022)Gabbard, Gillis, Chatelain, and M.~van
  Rees}]{Gabbard2021}
\bibinfo{author}{J.~Gabbard}, \bibinfo{author}{T.~Gillis},
  \bibinfo{author}{P.~Chatelain}, \bibinfo{author}{W.~M.~van Rees},
\newblock \bibinfo{title}{{An immersed interface method for the 2D
  vorticity-velocity Navier-Stokes equations with multiple bodies}},
\newblock \bibinfo{journal}{Journal of Computational Physics}
  \bibinfo{volume}{464} (\bibinfo{year}{2022}).
\bibitem[{Angot et~al.(1999)Angot, Bruneau, and Fabrie}]{Angot:1999}
\bibinfo{author}{P.~Angot}, \bibinfo{author}{C.-H. Bruneau},
  \bibinfo{author}{P.~Fabrie},
\newblock \bibinfo{title}{A penalization method to take into account obstacles
  in incompressible viscous flows},
\newblock \bibinfo{journal}{Numerische Mathematik} \bibinfo{volume}{81}
  (\bibinfo{year}{1999}) \bibinfo{pages}{497--520}.
\bibitem[{Coquerelle and Cottet(2008)}]{Coquerelle:2008}
\bibinfo{author}{M.~Coquerelle}, \bibinfo{author}{G.-H. Cottet},
\newblock \bibinfo{title}{A vortex level set method for the two-way coupling of
  an incompressible fluid with colliding rigid bodies},
\newblock \bibinfo{journal}{Journal of Computational Physics}
  \bibinfo{volume}{227} (\bibinfo{year}{2008}) \bibinfo{pages}{9121--9137}.
\bibitem[{Rossinelli et~al.(2010)Rossinelli, Bergdorf, Cottet, and
  Koumoutsakos}]{Rossinelli2010}
\bibinfo{author}{D.~Rossinelli}, \bibinfo{author}{M.~Bergdorf},
  \bibinfo{author}{G.-H. Cottet}, \bibinfo{author}{P.~Koumoutsakos},
\newblock \bibinfo{title}{{GPU accelerated simulations of bluff body flows
  using vortex particle methods}},
\newblock \bibinfo{journal}{Journal of Computational Physics}
  \bibinfo{volume}{229} (\bibinfo{year}{2010}) \bibinfo{pages}{3316--3333}.
\bibitem[{Gazzola et~al.(2011)Gazzola, Chatelain, van Rees, and
  Koumoutsakos}]{Gazzola:2011a}
\bibinfo{author}{M.~Gazzola}, \bibinfo{author}{P.~Chatelain},
  \bibinfo{author}{W.~M. van Rees}, \bibinfo{author}{P.~Koumoutsakos},
\newblock \bibinfo{title}{Simulations of single and multiple swimmers with
  non-divergence free deforming geometries},
\newblock \bibinfo{journal}{Journal of Computational Physics}
  \bibinfo{volume}{230} (\bibinfo{year}{2011}) \bibinfo{pages}{7093--7114}.
\bibitem[{Rossinelli et~al.(2015)Rossinelli, Hejazialhosseini, van Rees,
  Gazzola, Bergdorf, and Koumoutsakos}]{Rossinelli2015}
\bibinfo{author}{D.~Rossinelli}, \bibinfo{author}{B.~Hejazialhosseini},
  \bibinfo{author}{W.~van Rees}, \bibinfo{author}{M.~Gazzola},
  \bibinfo{author}{M.~Bergdorf}, \bibinfo{author}{P.~Koumoutsakos},
\newblock \bibinfo{title}{{MRAG-I2D: Multi-resolution adapted grids for
  remeshed vortex methods on multicore architectures}},
\newblock \bibinfo{journal}{Journal of Computational Physics}
  \bibinfo{volume}{288} (\bibinfo{year}{2015}) \bibinfo{pages}{1--18}.
\bibitem[{Gillis et~al.(2018)Gillis, Winckelmans, and Chatelain}]{Gillis2018}
\bibinfo{author}{T.~Gillis}, \bibinfo{author}{G.~Winckelmans},
  \bibinfo{author}{P.~Chatelain},
\newblock \bibinfo{title}{{Fast immersed interface Poisson solver for 3D
  unbounded problems around arbitrary geometries}},
\newblock \bibinfo{journal}{Journal of Computational Physics}
  \bibinfo{volume}{354} (\bibinfo{year}{2018}) \bibinfo{pages}{403 -- 416}.
\bibitem[{Bernier et~al.(2019)Bernier, Gazzola, Ronsse, and
  Chatelain}]{Bernier2019}
\bibinfo{author}{C.~Bernier}, \bibinfo{author}{M.~Gazzola},
  \bibinfo{author}{R.~Ronsse}, \bibinfo{author}{P.~Chatelain},
\newblock \bibinfo{title}{{Simulations of propelling and energy harvesting
  articulated bodies via vortex particle-mesh methods}},
\newblock \bibinfo{journal}{Journal of Computational Physics}
  \bibinfo{volume}{392} (\bibinfo{year}{2019}) \bibinfo{pages}{34--55}.
\bibitem[{Bhosale et~al.(2021)Bhosale, Parthasarathy, and
  Gazzola}]{Bhosale2021}
\bibinfo{author}{Y.~Bhosale}, \bibinfo{author}{T.~Parthasarathy},
  \bibinfo{author}{M.~Gazzola},
\newblock \bibinfo{title}{A remeshed vortex method for mixed rigid/soft body
  fluid--structure interaction},
\newblock \bibinfo{journal}{Journal of Computational Physics}
  \bibinfo{volume}{444} (\bibinfo{year}{2021}) \bibinfo{pages}{110577}.
\bibitem[{Engels et~al.(2015)Engels, Kolomenskiy, Schneider, and
  Sesterhenn}]{engels2015numerical}
\bibinfo{author}{T.~Engels}, \bibinfo{author}{D.~Kolomenskiy},
  \bibinfo{author}{K.~Schneider}, \bibinfo{author}{J.~Sesterhenn},
\newblock \bibinfo{title}{Numerical simulation of fluid--structure interaction
  with the volume penalization method},
\newblock \bibinfo{journal}{Journal of Computational Physics}
  \bibinfo{volume}{281} (\bibinfo{year}{2015}) \bibinfo{pages}{96--115}.
\bibitem[{Poncet(2009)}]{Poncet2009}
\bibinfo{author}{P.~Poncet},
\newblock \bibinfo{title}{{Analysis of an immersed boundary method for
  three-dimensional flows in vorticity formulation}},
\newblock \bibinfo{journal}{Journal of Computational Physics}
  \bibinfo{volume}{228} (\bibinfo{year}{2009}) \bibinfo{pages}{7268--7288}.
\bibitem[{Soltani and Ghomizad(2016)}]{Soltani2016}
\bibinfo{author}{E.~Soltani}, \bibinfo{author}{M.~B. Ghomizad},
\newblock \bibinfo{title}{Immersed boundary method for the simulation of heat
  transfer and fluid flow based on vorticity–velocity formulation},
\newblock \bibinfo{journal}{Numerical Heat Transfer, Part B: Fundamentals}
  \bibinfo{volume}{70} (\bibinfo{year}{2016}) \bibinfo{pages}{25--46}.
\bibitem[{Bourantas et~al.(2023)Bourantas, Zwick, Lavier, Loukopoulos, Dimas,
  Wittek, and Miller}]{bourantas2023immersed}
\bibinfo{author}{G.~C. Bourantas}, \bibinfo{author}{B.~F. Zwick},
  \bibinfo{author}{T.~P. Lavier}, \bibinfo{author}{V.~C. Loukopoulos},
  \bibinfo{author}{A.~A. Dimas}, \bibinfo{author}{A.~Wittek},
  \bibinfo{author}{K.~Miller},
\newblock \bibinfo{title}{An immersed boundary vector potential-vorticity
  meshless solver of the incompressible navier--stokes equation},
\newblock \bibinfo{journal}{International Journal for Numerical Methods in
  Fluids} \bibinfo{volume}{95} (\bibinfo{year}{2023})
  \bibinfo{pages}{143--175}.
\bibitem[{Wang and Eldredge(2015)}]{WandE2015}
\bibinfo{author}{C.~Wang}, \bibinfo{author}{J.~D. Eldredge},
\newblock \bibinfo{title}{{Strongly coupled dynamics of fluids and rigid-body
  systems with the immersed boundary projection method}},
\newblock \bibinfo{journal}{Journal of Computational Physics}
  \bibinfo{volume}{295} (\bibinfo{year}{2015}) \bibinfo{pages}{87--113}.
\bibitem[{Udaykumar et~al.(1999)Udaykumar, Mittal, and Shyy}]{Udaykumar1999}
\bibinfo{author}{H.~Udaykumar}, \bibinfo{author}{R.~Mittal},
  \bibinfo{author}{W.~Shyy},
\newblock \bibinfo{title}{Computation of solid--liquid phase fronts in the
  sharp interface limit on fixed grids},
\newblock \bibinfo{journal}{Journal of computational physics}
  \bibinfo{volume}{153} (\bibinfo{year}{1999}) \bibinfo{pages}{535--574}.
\bibitem[{Li(1997)}]{Li1997}
\bibinfo{author}{Z.~Li},
\newblock \bibinfo{title}{Immersed interface methods for moving interface
  problems},
\newblock \bibinfo{journal}{Numerical algorithms} \bibinfo{volume}{14}
  (\bibinfo{year}{1997}) \bibinfo{pages}{269--293}.
\bibitem[{Brehm and Fasel(2011)}]{Brehm2011}
\bibinfo{author}{C.~Brehm}, \bibinfo{author}{H.~Fasel},
\newblock \bibinfo{title}{Immersed interface method for solving the
  incompressible navier-stokes equations with moving boundaries},
\newblock in: \bibinfo{booktitle}{49th AIAA Aerospace Sciences Meeting
  including the New Horizons Forum and Aerospace Exposition},
  \bibinfo{year}{2011}, p. \bibinfo{pages}{758}.
\bibitem[{Xu and Wang(2006)}]{Xu2006}
\bibinfo{author}{S.~Xu}, \bibinfo{author}{Z.~J. Wang},
\newblock \bibinfo{title}{An immersed interface method for simulating the
  interaction of a fluid with moving boundaries},
\newblock \bibinfo{journal}{Journal of Computational Physics}
  \bibinfo{volume}{216} (\bibinfo{year}{2006}) \bibinfo{pages}{454--493}.
\bibitem[{Udaykumar et~al.(2001)Udaykumar, Mittal, Rampunggoon, and
  Khanna}]{Udaykumar2001}
\bibinfo{author}{H.~Udaykumar}, \bibinfo{author}{R.~Mittal},
  \bibinfo{author}{P.~Rampunggoon}, \bibinfo{author}{A.~Khanna},
\newblock \bibinfo{title}{A sharp interface cartesian grid method for
  simulating flows with complex moving boundaries},
\newblock \bibinfo{journal}{Journal of computational physics}
  \bibinfo{volume}{174} (\bibinfo{year}{2001}) \bibinfo{pages}{345--380}.
\bibitem[{Mittal et~al.(2008)Mittal, Dong, Bozkurttas, Najjar, Vargas, and
  Von~Loebbecke}]{Mittal2008}
\bibinfo{author}{R.~Mittal}, \bibinfo{author}{H.~Dong},
  \bibinfo{author}{M.~Bozkurttas}, \bibinfo{author}{F.~Najjar},
  \bibinfo{author}{A.~Vargas}, \bibinfo{author}{A.~Von~Loebbecke},
\newblock \bibinfo{title}{A versatile sharp interface immersed boundary method
  for incompressible flows with complex boundaries},
\newblock \bibinfo{journal}{Journal of computational physics}
  \bibinfo{volume}{227} (\bibinfo{year}{2008}) \bibinfo{pages}{4825--4852}.
\bibitem[{Brehm et~al.(2019)Brehm, Barad, and Kiris}]{Brehm2019}
\bibinfo{author}{C.~Brehm}, \bibinfo{author}{M.~F. Barad},
  \bibinfo{author}{C.~C. Kiris},
\newblock \bibinfo{title}{Development of immersed boundary computational
  aeroacoustic prediction capabilities for open-rotor noise},
\newblock \bibinfo{journal}{Journal of Computational Physics}
  \bibinfo{volume}{388} (\bibinfo{year}{2019}) \bibinfo{pages}{690--716}.
\bibitem[{Boustani et~al.(2021)Boustani, Barad, Kiris, and
  Brehm}]{boustani2021immersed}
\bibinfo{author}{J.~Boustani}, \bibinfo{author}{M.~F. Barad},
  \bibinfo{author}{C.~C. Kiris}, \bibinfo{author}{C.~Brehm},
\newblock \bibinfo{title}{An immersed boundary fluid--structure interaction
  method for thin, highly compliant shell structures},
\newblock \bibinfo{journal}{Journal of Computational Physics}
  \bibinfo{volume}{438} (\bibinfo{year}{2021}) \bibinfo{pages}{110369}.
\bibitem[{Yang and Balaras(2006)}]{Yang2006}
\bibinfo{author}{J.~Yang}, \bibinfo{author}{E.~Balaras},
\newblock \bibinfo{title}{An embedded-boundary formulation for large-eddy
  simulation of turbulent flows interacting with moving boundaries},
\newblock \bibinfo{journal}{Journal of Computational Physics}
  \bibinfo{volume}{215} (\bibinfo{year}{2006}) \bibinfo{pages}{12--40}.
\bibitem[{Seo and Mittal(2011)}]{Seo2011}
\bibinfo{author}{J.~H. Seo}, \bibinfo{author}{R.~Mittal},
\newblock \bibinfo{title}{A sharp-interface immersed boundary method with
  improved mass conservation and reduced spurious pressure oscillations},
\newblock \bibinfo{journal}{Journal of computational physics}
  \bibinfo{volume}{230} (\bibinfo{year}{2011}) \bibinfo{pages}{7347--7363}.
\bibitem[{Gabbard and van Rees(2023)}]{gabbard2023high}
\bibinfo{author}{J.~Gabbard}, \bibinfo{author}{W.~M. van Rees},
\newblock \bibinfo{title}{A high-order 3d immersed interface finite difference
  method for the advection-diffusion equation},
\newblock in: \bibinfo{booktitle}{AIAA SCITECH 2023 Forum},
  \bibinfo{year}{2023}, p. \bibinfo{pages}{2480}.
\bibitem[{Williamson(1980)}]{Williamson1980}
\bibinfo{author}{J.~H. Williamson},
\newblock \bibinfo{title}{{Low-storage Runge-Kutta schemes}},
\newblock \bibinfo{journal}{Journal of Computational Physics}
  \bibinfo{volume}{35} (\bibinfo{year}{1980}) \bibinfo{pages}{48--56}.
\bibitem[{Quartapelle(1993)}]{Quartapelle1993}
\bibinfo{author}{L.~Quartapelle}, \bibinfo{title}{Numerical Solution of the
  Incompressible Navier-Stokes Equations}, \bibinfo{publisher}{Birkh{\"{a}}user
  Basel}, \bibinfo{year}{1993}.
\bibitem[{Noca(1997)}]{Noca1997}
\bibinfo{author}{F.~Noca}, \bibinfo{title}{{On the evaluation of time-dependent
  fluid-dynamic forces on bluff bodies}}, Ph.D. thesis, California Institute of
  Technology, \bibinfo{year}{1997}.
\bibitem[{Shen et~al.(2009)Shen, Chan, and Lin}]{shen_calculation_2009}
\bibinfo{author}{L.~Shen}, \bibinfo{author}{E.-S. Chan},
  \bibinfo{author}{P.~Lin},
\newblock \bibinfo{title}{Calculation of hydrodynamic forces acting on a
  submerged moving object using immersed boundary method},
\newblock \bibinfo{journal}{Computers \& Fluids} \bibinfo{volume}{38}
  (\bibinfo{year}{2009}) \bibinfo{pages}{691--702}.
\bibitem[{Bergmann and Iollo(2016)}]{bergmann_bioinspired_2016}
\bibinfo{author}{M.~Bergmann}, \bibinfo{author}{A.~Iollo},
\newblock \bibinfo{title}{Bioinspired swimming simulations},
\newblock \bibinfo{journal}{Journal of Computational Physics}
  \bibinfo{volume}{323} (\bibinfo{year}{2016}) \bibinfo{pages}{310--321}.
\bibitem[{Nangia et~al.(2017)Nangia, Johansen, Patankar, and
  Bhalla}]{nangia_moving_2017}
\bibinfo{author}{N.~Nangia}, \bibinfo{author}{H.~Johansen},
  \bibinfo{author}{N.~A. Patankar}, \bibinfo{author}{A.~P.~S. Bhalla},
\newblock \bibinfo{title}{A moving control volume approach to computing
  hydrodynamic forces and torques on immersed bodies},
\newblock \bibinfo{journal}{Journal of Computational Physics}
  \bibinfo{volume}{347} (\bibinfo{year}{2017}) \bibinfo{pages}{437--462}.
\bibitem[{Towers(2009)}]{Towers2009}
\bibinfo{author}{J.~D. Towers},
\newblock \bibinfo{title}{Finite difference methods for approximating heaviside
  functions},
\newblock \bibinfo{journal}{Journal of Computational Physics}
  \bibinfo{volume}{228} (\bibinfo{year}{2009}) \bibinfo{pages}{3478--3489}.
\bibitem[{Gillis(2019)}]{GillisThesis}
\bibinfo{author}{T.~Gillis}, \bibinfo{title}{{Accurate and efficient treatment
  of solid boundaries for the vortex particle-mesh method}}, Ph.D. thesis,
  UCLouvain, \bibinfo{year}{2019}.
\bibitem[{Koumoutsakos and Leonard(1995)}]{PandL1995}
\bibinfo{author}{P.~Koumoutsakos}, \bibinfo{author}{A.~Leonard},
\newblock \bibinfo{title}{High-resolution simulations of the flow around an
  impulsively started cylinder using vortex methods},
\newblock \bibinfo{journal}{Journal of Fluid Mechanics} \bibinfo{volume}{296}
  (\bibinfo{year}{1995}) \bibinfo{pages}{1--38}.
\bibitem[{Eldredge and Wang(2010)}]{EandW2010}
\bibinfo{author}{J.~D. Eldredge}, \bibinfo{author}{C.~Wang},
\newblock \bibinfo{title}{High-fidelity simulations and low-order modeling of a
  rapidly pitching plate},
\newblock \bibinfo{journal}{40th Fluid Dynamics Conference and Exhibit}
  (\bibinfo{year}{2010}).
\bibitem[{Schneiders et~al.(2013)Schneiders, Hartmann, Meinke, and
  Schröder}]{Schneiders2013}
\bibinfo{author}{L.~Schneiders}, \bibinfo{author}{D.~Hartmann},
  \bibinfo{author}{M.~Meinke}, \bibinfo{author}{W.~Schröder},
\newblock \bibinfo{title}{{An accurate moving boundary formulation in cut-cell
  methods}},
\newblock \bibinfo{journal}{Journal of Computational Physics}
  \bibinfo{volume}{235} (\bibinfo{year}{2013}) \bibinfo{pages}{786–809}.
\bibitem[{Chatelain et~al.(2013)Chatelain, Backaert, Winckelmans, and
  Kern}]{Chatelain2013}
\bibinfo{author}{P.~Chatelain}, \bibinfo{author}{S.~Backaert},
  \bibinfo{author}{G.~Winckelmans}, \bibinfo{author}{S.~Kern},
\newblock \bibinfo{title}{Large eddy simulation of wind turbine wakes},
\newblock \bibinfo{journal}{Flow, turbulence and combustion}
  \bibinfo{volume}{91} (\bibinfo{year}{2013}) \bibinfo{pages}{587--605}.
\bibitem[{Guilmineau and Queutey(2002)}]{Guilmineau2002}
\bibinfo{author}{E.~Guilmineau}, \bibinfo{author}{P.~Queutey},
\newblock \bibinfo{title}{{A numerical simulation of vortex shedding from an
  oscillating circular cylinder}},
\newblock \bibinfo{journal}{Journal of Fluids and Structures}
  \bibinfo{volume}{16(6)} (\bibinfo{year}{2002}) \bibinfo{pages}{773--794}.
\bibitem[{Yang et~al.(2009)Yang, Zhang, Li, and He}]{Yang2009}
\bibinfo{author}{X.~Yang}, \bibinfo{author}{X.~Zhang}, \bibinfo{author}{Z.~Li},
  \bibinfo{author}{G.-W. He},
\newblock \bibinfo{title}{{A smoothing technique for discrete delta functions
  with application to immersed boundary method in moving boundary
  simulations}},
\newblock \bibinfo{journal}{Journal of Computational Physics}
  \bibinfo{volume}{228(20)} (\bibinfo{year}{2009}) \bibinfo{pages}{7821--7836}.
\bibitem[{Namkoong et~al.(2008)Namkoong, Yoo, and Choi}]{Namkoong2008}
\bibinfo{author}{K.~Namkoong}, \bibinfo{author}{J.~Y. Yoo},
  \bibinfo{author}{H.~G. Choi},
\newblock \bibinfo{title}{{Numerical analysis of two-dimensional motion of a
  freely falling circular cylinder in an infinite fluid}},
\newblock \bibinfo{journal}{Journal of Fluid Mechanics} \bibinfo{volume}{604}
  (\bibinfo{year}{2008}) \bibinfo{pages}{33--53}.
\bibitem[{Ern et~al.(2012)Ern, Risso, Fabre, and Magnaudet}]{Ern2012}
\bibinfo{author}{P.~E. Ern}, \bibinfo{author}{F.~Risso},
  \bibinfo{author}{D.~Fabre}, \bibinfo{author}{J.~Magnaudet},
\newblock \bibinfo{title}{{Wake-Induced oscillatory paths of bodies freely
  rising or falling in fluids}},
\newblock \bibinfo{journal}{Annual Review of Fluid Mechanics}
  \bibinfo{volume}{44} (\bibinfo{year}{2012}) \bibinfo{pages}{97–121}.
\bibitem[{Ahn and Kallinderis(2006)}]{Ahn2006}
\bibinfo{author}{H.~T. Ahn}, \bibinfo{author}{Y.~Kallinderis},
\newblock \bibinfo{title}{{Strongly coupled flow/structure interactions with a
  geometrically conservative ALE scheme on general hybrid meshes}},
\newblock \bibinfo{journal}{Journal of Computational Physics}
  \bibinfo{volume}{219} (\bibinfo{year}{2006}) \bibinfo{pages}{671–696}.
\bibitem[{Bao et~al.(2012)Bao, Huang, Zhou, and Tu}]{Bao2012}
\bibinfo{author}{Y.~Bao}, \bibinfo{author}{C.~Huang},
  \bibinfo{author}{D.~Zhou}, \bibinfo{author}{Z.~Tu, Jiahuang and.~Han},
\newblock \bibinfo{title}{{Two-degree-of-freedom flow-induced vibrations on
  isolated and tandem cylinders with varying natural frequency ratios}},
\newblock \bibinfo{journal}{Journal of Fluids and Structures}
  \bibinfo{volume}{35} (\bibinfo{year}{2012}) \bibinfo{pages}{50--75}.
\bibitem[{Borazjani and Sotiropoulos(2009)}]{BandS2009}
\bibinfo{author}{I.~Borazjani}, \bibinfo{author}{F.~Sotiropoulos},
\newblock \bibinfo{title}{{Vortex-induced vibrations of two cylinders in tandem
  arrangement in the proximity–wake interference region}},
\newblock \bibinfo{journal}{Journal of Fluid Mechanics} \bibinfo{volume}{621}
  (\bibinfo{year}{2009}) \bibinfo{pages}{321--364}.
\bibitem[{Gillis and M.~van Rees(2022)}]{Gillis:2022}
\bibinfo{author}{T.~Gillis}, \bibinfo{author}{W.~M.~van Rees},
\newblock \bibinfo{title}{{MURPHY---A Scalable Multiresolution Framework for
  Scientific Computing on 3D Block-Structured Collocated Grids}},
\newblock \bibinfo{journal}{SIAM Journal on Scientific Computing}
  \bibinfo{volume}{44} (\bibinfo{year}{2022}).

\end{thebibliography}

\end{document}